%% file: main.tex
\documentclass[12pt]{article}
\usepackage[a4paper]{geometry}
\usepackage{graphicx, caption}
\usepackage{setspace}
\usepackage{bm}
\usepackage{amsmath}
\usepackage{siunitx}
\usepackage{caption}
\usepackage{subcaption}
\usepackage[style=nature]{biblatex}
\usepackage[font=scriptsize,skip=2pt]{caption}
\usepackage{authblk}
\usepackage{subfiles}
\usepackage{subcaption}

\title{Initial Fulcher band observations from high resolution spectroscopy in the MAST-U divertor}
\author[1]{N. Osborne}
\author[2]{K. Verhaegh}
\author[3]{M.D. Bowden}
\author[4]{T. Wijkamp}
\author[5]{N. Lonigro}
\author[6]{P. Ryan}
\author[7]{E. Pawelec}
\author[8]{B. Lipschultz}
\author[9]{V. Soukhanovskii}
\author[10]{T. van den Biggelaar}
\author[11]{the MAST-U team}
\affil[1,3] {University of Liverpool}
\affil[1,2,5,6,11]{Culham Centre for Fusion Energy, Culham, United Kingdom}
\affil[4,10]{Eindhoven University of Technology, Eindhoven, Netherlands}
\affil[7]{University of Opole, Opole, Poland}
\affil[5,8]{University of York, York, United Kingdom}
\affil[9]{Lawrence Livermore National Laboratory, Livermore, United States}

\begin{document}
\date{}
\maketitle

\begin{abstract}
	High resolution $\text{D}_{2}$ Fulcher band spectroscopy was used in the MAST-U divertors during Super-X and elongated conventional divertor density ramps with $\text{D}_{2}$ fuelling from the mid-plane high-field side.  In the Super-X case (density ramp from Greenwald fraction 0.12 to 0.24), the upper divertor showed ground state rotational temperatures of the $\text{D}_{2}$ molecules increasing from $\sim$6000 K, starting at the detachment onset, to $\sim$9000 K during deepening detachment.  This was correlated with the movement of the Fulcher emission region towards the X-point, which is in turn correlated with the ionisation source. The increase in rotational temperature occured throughout the divertor except near the divertor entrance, where ionisation was still the dominant process. Qualitative agreement was obtained between the lower and upper divertor. Similar rotational temperatures were obtained in the elongated divertor before the detachment onset, although the increase in rotational temperature during detachment was less clearly observed as less deep detachment was obtained. 
 
 The measured vibrational distribution of the upper Fulcher state (first four bands) does not agree with a ground state Boltzmann distribution but shows a characteristically elevated population in the $\nu = 2$ and $\nu = 3$ bands in particular.  The populations of the $\nu = 2$ and $\nu = 3$ band relative to the $\nu = 0$ band are strongly correlated to the $\textit{rotational}$ temperature.
\end{abstract}

\abovedisplayskip=0pt
\belowdisplayskip=0pt

\subfile{Sections/1.Intro}
\subfile{Sections/2.Method_discharge_info}

\subfile{Sections/3.Results_rotational}

\subfile{Sections/4.Results_vibrational}
\subfile{Sections/5.Discussion}
\subfile{Sections/6.Summary}

\subfile{Sections/Appendix}

\clearpage
\printbibliography

\end{document}

%% file: Sections/1.Intro.tex
\section{Introduction}\label{sec1}

One of the major issues facing future tokamak devices is that the heat fluxes arriving at the divertor target plates during operation can exceed material limits \cite{Pitts2019,Wenninger2014} by orders of magnitude unless steps are taken to ensure otherwise.  Divertors in reactors will need to operate in a so-called ``detached" state in which simultaneous particle (e.g. ion), momentum and power losses occur \cite{Verhaegh2019AnDetachment,Verhaegh2021AConditions,Lipschultz1998,Stangeby2018,Krasheninnikov2017}. This facilitates reductions in the ion target flux, as the divertor plasma is cooled, leading to orders of magnitude heat flux reductions. \\

\begin{figure}[th]
  \centering
  \includegraphics[width=0.8\linewidth]{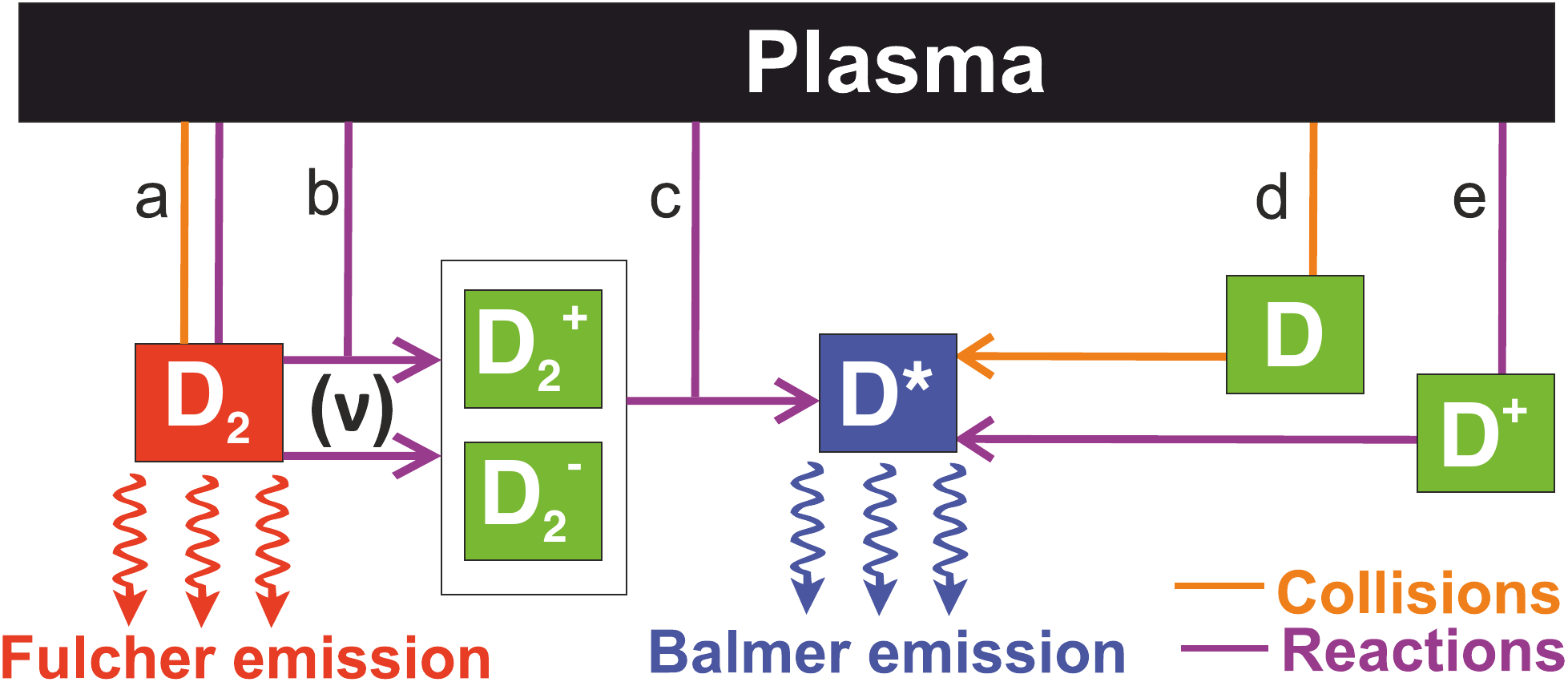}
 \vspace*{0.3cm}
  \caption{Adapted from \cite{Verhaegh2021a}. Schematic overview of various reactions and collisions resulting in excited atoms and molecules and thus hydrogen Balmer emission and $D_2$ Fulcher emission. a-e indicate different collision/reaction mechanisms and are referred to in the text.}
  \label{fig: schem_fulcher_balmer}
\end{figure}

Collisions and reactions between the plasma and the molecules in the divertor volume can result in rovibronic (e.g. rotational, vibrational and electronic) excitation. Some of these collisions lead to momentum and energy transfer from the plasma to the neutral molecular cloud, resulting in power and momentum dissipation especially at the `killer' flux tubes \cite{Moulton2017,Myatra} (figure \ref{fig: schem_fulcher_balmer} a). Vibrationally excited molecules can undergo plasma-chemistry reactions with the plasma, resulting in molecular ions such as $D_2^+$ and, potentially, $D_2^- \rightarrow D^- + D$ (figure \ref{fig: schem_fulcher_balmer} b). These ions react with the plasma (figure \ref{fig: schem_fulcher_balmer} c), resulting in additional dissociative channels leading to a hydrogen atom source (Molecular Activated Dissociation - MAD) as well as ion sources and sinks through Molecular Activated Ionisation (MAI) and Molecular Activated Recombination (MAR) \cite{Krasheninnikov1996PlasmaDetachment,Verhaegh2021AConditions,Verhaegh2022SpectroscopicDivertor,Verhaegh2023b}. Recent spectroscopic analysis has indicated that MAR and MAD can influence the progression of detachment significantly on MAST Upgrade \cite{Verhaegh2022SpectroscopicDivertor}, TCV \cite{Verhaegh2021AConditions,Perek2022} and JET \cite{Karhunen2022a}\\

\subsection*{MAST Upgrade and the Super-X divertor} 

It is currently unknown whether conventional divertors will facilitate sufficient power exhaust performance whilst maintaining acceptable core performance \cite{Wenninger2014}. Therefore, as a risk mitigation strategy, alternative divertor concepts (ADC) are being developed and tested \cite{Havlickova2015,Theiler2017,Moulton2023,Verhaegh2022SpectroscopicDivertor,Kotschenreuther2013,Verhaegh2023b}. The Mega Amp Spherical Tokamak Upgrade (MAST-U) is a new tokamak, based at UKAEA in Culham, UK, that supports a range of ADCs in tightly baffled upper and lower divertor chambers, including the Super-X divertor, as well as a more conventional divertor \cite{Havlickova2015,Moulton2017,Moulton2023,Verhaegh2022SpectroscopicDivertor,Verhaegh2023b}. In the Super-X divertor, the strike point is increased to a larger target radius, resulting in lower magnetic field at the target and a high total flux expansion \cite{Lipschultz2016,Cowley2022}. This is predicted to enhance access to plasma detachment and optimise power exhaust capabilities.\\

The first MAST-U results have shown a greatly improved power exhaust performance in the Super-X divertor \cite{Moulton2023,Verhaegh2022SpectroscopicDivertor}. Detailed analysis of the physics mechanisms during detachment in the Super-X divertor have shown that plasma-molecular chemistry, including MAR and MAD, plays an unprecedentedly strong role in the tightly baffled divertor chambers when the ionisation source is significantly detached from the target.  This results in a build-up of MAR downstream of the ionisation region towards the target \cite{Verhaegh2022SpectroscopicDivertor,Verhaegh2023b}. \\

\subsection*{Visible spectroscopy: $\text{D}^*$ Balmer and $\text{D}_{2}$ Fulcher emission}
Plasma-neutral interactions result in excited hydrogen atoms and molecules that can be detected through visible spectroscopy of both hydrogen Balmer ($D^*$) and Fulcher ($D_2$) emission. Analysing such information can provide quantitative information on the divertor conditions \cite{Verhaegh2021,Verhaegh2019a,Verhaegh2017,Fantz1998SpectroscopicMolecules,Perek2022,Wijkamp2023,Bowman2020,Lomanowski2020,Hollmann2006SpectroscopicEdge} as well as the plasma-neutral interaction processes, providing critical information for diagnosing and developing an understanding of power exhaust and detachment in tokamak divertors. \\

Excited hydrogen atoms arise from atomic interactions, such as electron impact excitation (figure \ref{fig: schem_fulcher_balmer} d) and electron-ion recombination (EIR) (figure \ref{fig: schem_fulcher_balmer} e), as well as plasma-molecular interactions that ultimately result in excited hydrogen atoms. In detached conditions, the latter possibility mostly arises from interactions between the plasma and molecular ions ($D_2^+$ and $D_2^- \rightarrow D^- + D$) (figure \ref{fig: schem_fulcher_balmer}). The development of new Balmer emission analysis techniques \cite{Verhaegh2019a,Verhaegh2021}, such as the BaSPMI technique (Balmer Spectroscopy for Plasma–Molecule Interaction) \cite{Verhaegh2021AConditions}, enables separation of the hydrogen Balmer line emission into its different contributors. This information can be used quantitatively to estimate the ion sources and sinks from both plasma-atom and plasma-molecular interactions (ionisation, EIR, MAR, MAI) as well as the hydrogenic radiative and net power losses.\\

Unlike $D^*$ Balmer emission, the less well-known $D_2$ Fulcher band originates not from hydrogen atoms, but from the electronic de-excitation of hydrogen molecules between the electronic triplet states: $d^3\Pi^-_u\xrightarrow{}a^3\Sigma^+_g$. The threshold energy required for the electronic excitation which facilitates this Fulcher transition is about 12 eV.  Strong $D_2$ Fulcher band emission, however, occurs at 4-5 eV where a much larger molecular density (which increases with decreasing temperature \cite{Stangeby2017,Verhaegh2022SpectroscopicDivertor}) is balanced optimally with there still being a large enough number of hot electrons in the tail of the electron Boltzmann distribution.  This coincides with both the electron-impact dissociation and the (atomic) ionisation regions \cite{Verhaegh2022SpectroscopicDivertor}.  In this way, the location of Fulcher emission can be used as a temperature constraint as well as a proxy for the ionisation source \cite{Verhaegh2022SpectroscopicDivertor,Wijkamp2023}, which can be useful for divertor detachment studies and divertor detachment control.\\

\subsubsection*{The importance of the rotational and vibrational $\text{D}_{2}$ distribution}
The Fulcher band, though, has yet more to offer since information regarding the rotational and vibrational distributions of hydrogen molecules is also encoded within its multitude of transitions, and can be extracted via high resolution spectroscopy. If the rotational and vibrational distribution measurements follow a Boltzmann distribution, they may be summarised with a rotational and vibrational temperature respectively \cite{Fantz2002,Hollmann2006SpectroscopicEdge,Brezinsek2005}.  The rotational and vibrational structure of the Fulcher band is described in detail in the appendix.\\

Collisions between the plasma and the molecules result in energy and momentum transfer from the plasma to the molecules, which can be a critical ingredient to detachment according to SOLPS-ITER simulations \cite{Moulton2017,Park2018,Myatra}.  Assuming that the rotational temperature is a proxy for the gas temperature of the molecules and thus a reasonable indicator for kinetic energy (see appendix), diagnosing the rotational distribution and inferring the rotational temperature can be utilised to observe signatures of energy transfer and may be used to diagnose such transfers as well as verify simulated predictions experimentally.\\

The vibrational distribution plays a crucial role in plasma chemistry. Dissociative attachment ($e^- + D_2 \rightarrow D_2^- \rightarrow D^- + D$), molecular charge exchange ($D^+ + D_2 \rightarrow D_2^+ + D$), $D_2$ ionisation ($e^- + D_2 \rightarrow D_2^+ + 2e^-$) and electron-impact dissociation ($D_2 + e^- \rightarrow D + D + e^-$) all have cross-sections with strong dependencies on the vibrational distribution.  For molecular charge exchange and dissociative attachment in particular, the dependency spans multiple orders of magnitude. Reactions that generate molecular ions that drive the experimentally observed plasma chemistry are, therefore, driven to a great extent by the vibrational distribution. Although vibrationally resolved plasma-edge simulations exist \cite{Wischmeier2005,Fantz2006}, they generally utilise `effective' rates for plasma-molecular interactions. Within such simulations, a vibrational model is implicitly assumed in the derivation of the `effective' rates, whilst transport and plasma-wall interactions are assumed to have a negligible impact on the vibrational distribution \cite{Kotov2007}. Additionally, even if vibrationally resolved simulations are utilised, the reactions included that result in vibrational (de)excitation, the cross-sections used for such reactions, as well as the impact of plasma-wall interactions on the vibrational distribution are all debated in literature \cite{Wischmeier2005,Verhaegh2023a,Verhaegh2022SpectroscopicDivertor,Fantz2006,Reiter2018,Chandra2023,Holm2022}. Validating the various models and assumptions is critical for ensuring plasma-chemistry reaction mechanisms are correctly included in plasma-edge models; which is important for extrapolating current knowledge to reactor-class devices through plasma-edge simulations.

\subsection*{This paper}

The aim of this paper is to present an analysis of the first high resolution $D_2$ Fulcher band data obtained from the MAST-U divertor chambers in the Super-X divertor, as well as more conventional divertor topologies, from the machine's first experimental campaign in 2021. This analysis provides measurements of the rotational and vibrational distribution of the molecules during experiments where the core density is scanned to study the onset of, and the physics mechanisms during, detachment.\\

In section \ref{sec2}, both the divertor monitoring spectroscopy available on MAST-U during the first campaign, and the discharges for which high resolution Fulcher band spectra were obtained, are described.  Details of the structure of the Fulcher band, the theory and the established methods for analysing rotational and vibrational distributions for diatomic molecules such as $D_2$ are described in the appendix.  The initial results from MAST-U are presented in section \ref{sec:results}, followed by a discussion of these results in section \ref{sec:discussion}; and conclusion in section \ref{summary}.

%% file: Sections/2.Method_discharge_info.tex
\section{Experimental setup}\label{sec2}

\subsection*{Discharges examined in this paper}

\begin{figure}[th]
  \centering
  \renewcommand{\arraystretch}{0.5} 
  \begin{tabular}{ c @{\quad} c }
    \includegraphics[width=20em]{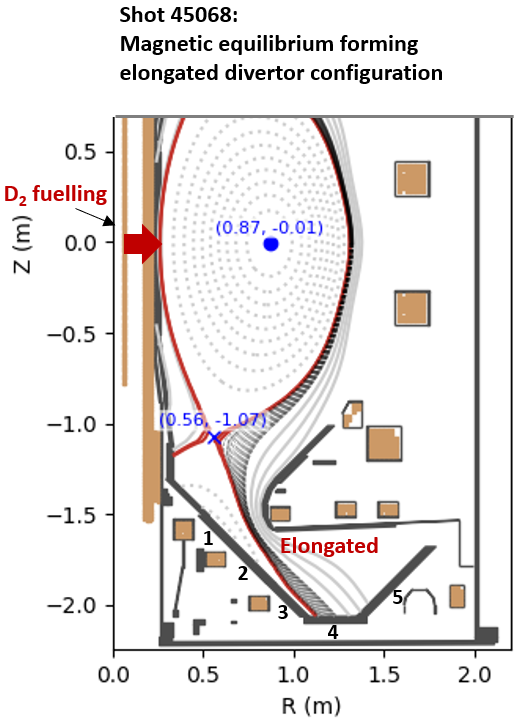} &
      \includegraphics[width=20em]{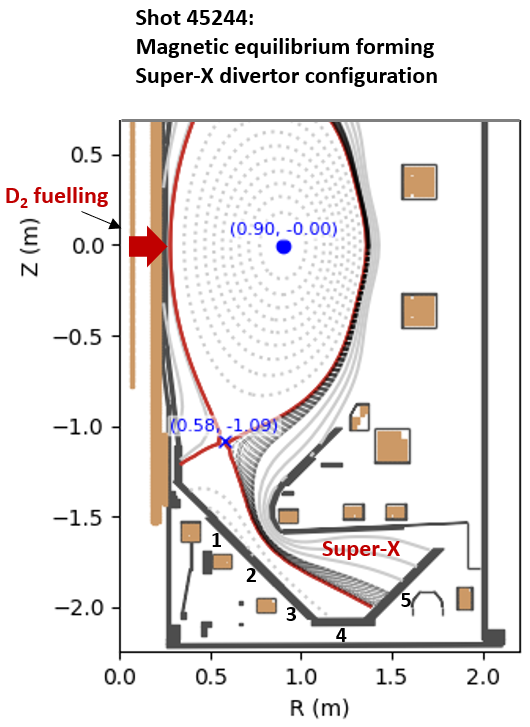} \\
    \scriptsize (a) &
      \scriptsize (b)
  \end{tabular}
  \vspace*{0.3cm}
  \caption{\it \footnotesize The magnetic equilbria forming (a) the elongated conventional divertor and (b) the Super-X divertor configurations for shots 45068 and 45244 respectively.  Note the numbering of the divertor tiles from 1 to 5.\label{equ}}
\end{figure}

Discharges 45068 and 45244 are both $I_p = 650$ kA Ohmic discharges with a fuelling ramp (from the mid-plane high-field side at the position indicated in figure \ref{equ}) to scan the core and separatrix densities and vary the degree of detachment. The magnetic equilibria are shown in figures \ref{equ} (a) and (b) respectively. The strike point for discharge 45244  in the Super-X configuration is on `tile 5' - see figure \ref{equ} (b).  The strike point for a conventional divertor discharge would be on `tile 3' (not shown), while that of discharge 45068 is on `tile 4' - see figure \ref{equ} (a).  For this reason discharge 45068 is referred to as an ``elongated conventional divertor" (ECD). This geometry has additional spatial coverage of the divertor leg with the various divertor spectrometers, in comparison to the ``conventional" divertor geometry (CD). \\

\begin{figure}[th]
  \centering
  \renewcommand{\arraystretch}{0.5} 
  \begin{tabular}{ c @{\quad} c }
    \includegraphics[width=18em]{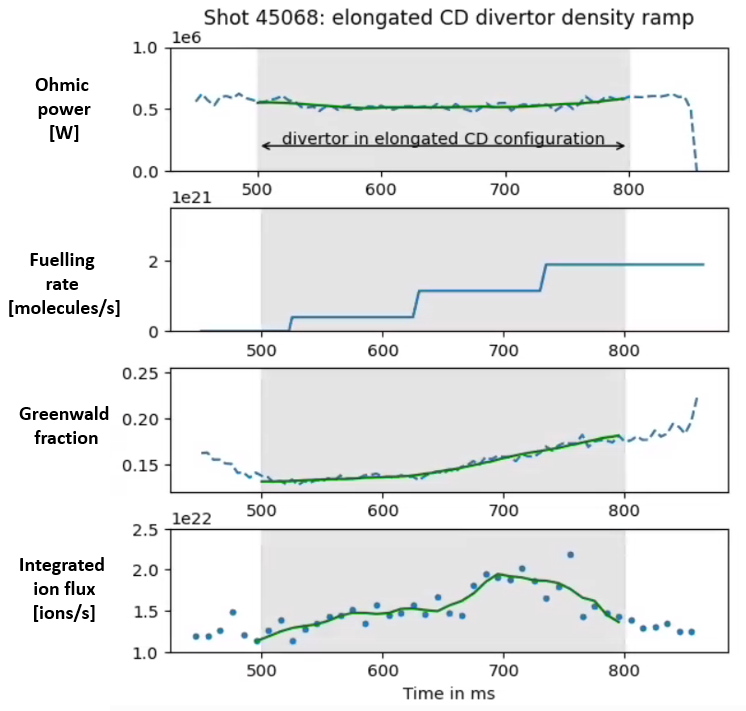} &
      \includegraphics[width=18em]{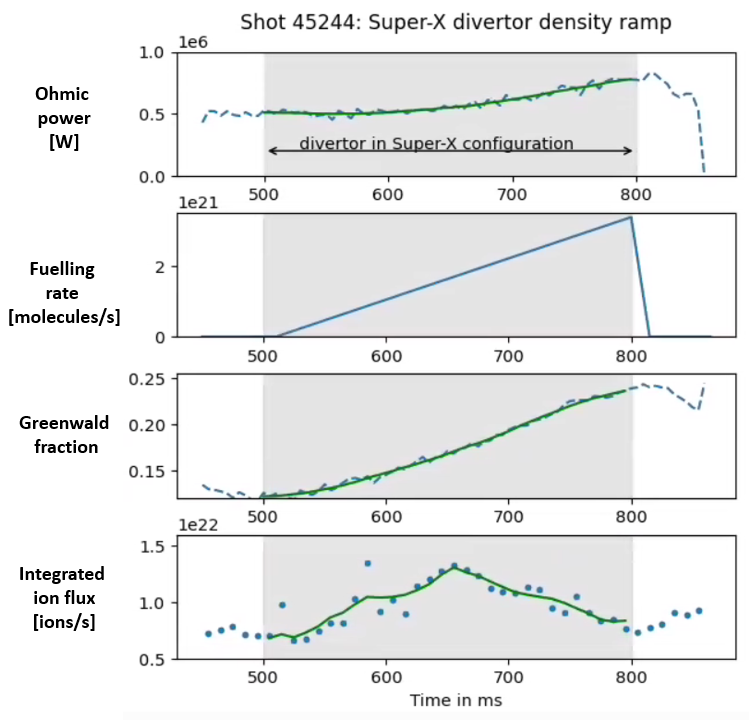} \\
    \scriptsize (a) &
      \scriptsize (b)
  \end{tabular}
  \vspace*{0.3cm}
  \caption{\it \footnotesize The Ohmic power, gas fuelling rate, Greenwald fraction and integrated ion flux (upper divertor) for: (a) Elongated conventional divertor shot 45068; and (b) Super-X divertor shot 45244.  In both cases fuelling is from the high-field side mid-plane.\label{one}}
\end{figure}

Details of the evolution of Ohmic power, $D_2$ fuelling gas flux, core line-averaged density (expressed as a fraction with respect to the Greenwald limit), and integrated ion flux (from Langmuir probe measurements) are shown in figures \ref{one} (a) and (b) for shots 45068 and 45244 respectively.  The shaded region in the graphs between 500 ms and 800 ms indicates the period during which the equilibria were fully formed and held constant.

\subsubsection*{High Resolution Divertor Spectroscopy in The MAST-U Divertor}

\begin{figure}[th]
  \centering
  \renewcommand{\arraystretch}{0.5} 
  \begin{tabular}{ c @{\quad} c }
    \includegraphics[width=20em]{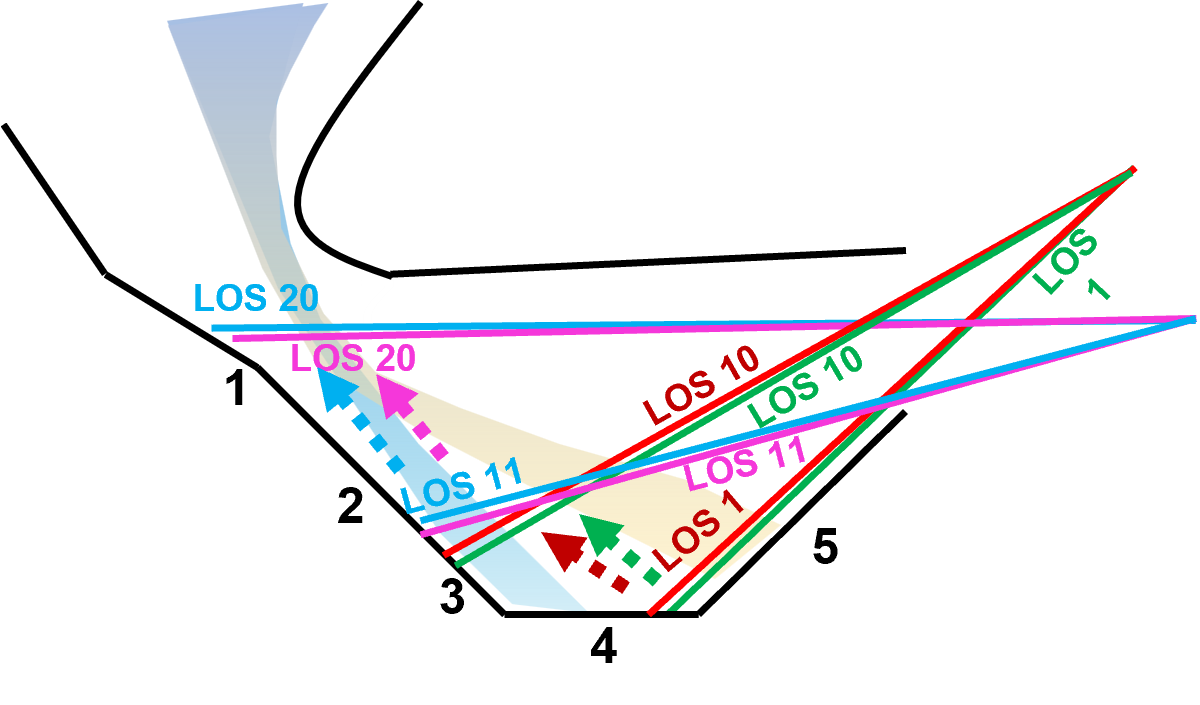} &
      \includegraphics[width=20em]{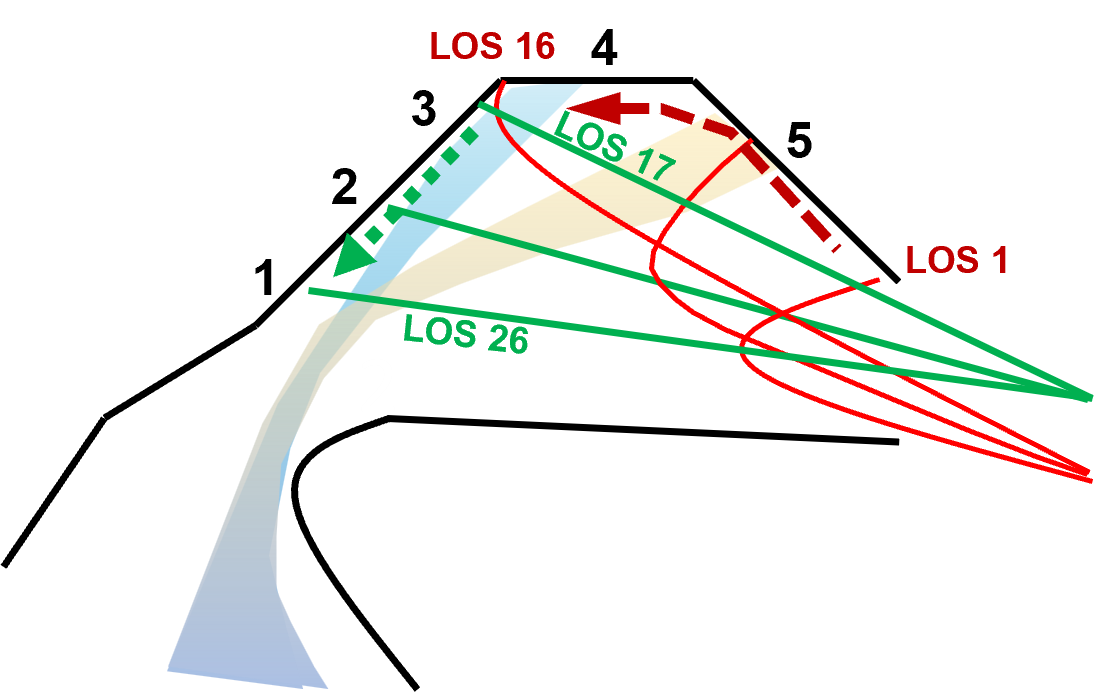} \\
    \scriptsize (a) &
      \scriptsize (b)
  \end{tabular}
  \vspace*{0.3cm}
  \caption{\it \footnotesize Depictions of the divertor spectroscopy.  (a) Shows the lower divertor in which the green and pink lines of sight are connected to the ``York" spectrometer; and the red and cyan to the ``CCFE" spectrometer.  (b) Shows the upper divertor in which all lines of sight are connected to the ``Dibs" spectrometer.  The black numbers refer to the tile numbers.  The cyan and orange shading indicate the elongated conventional divertor and Super-X divertor magnetic geometries respectively.\label{Ds}}
\end{figure}

During the first experimental campaign, three spectrometers were used for spectroscopy in the divertors \cite{Verhaegh2022SpectroscopicDivertor,Verhaegh2023b}.  The ``York" spectrometer and the ``CCFE" spectrometer covered 20 lines of sight each which were interspersed through the lower divertor as shown in figure \ref{Ds} (a).  Meanwhile, the ``Dibs" spectrometer surveyed the upper divertor with 16 lines of sight which have a tangential component and track across tiles 4 and 5;  and a further ten lines of sight fanning across tiles 1 to 3 in the poloidal plane as shown in figure \ref{Ds} (b).  For high resolution Fulcher band spectroscopy, the following gratings were used at the following wavelengths:  The ``York" spectrometer used a 1800 l/mm grating (with a spectral resolution of 0.06 nm) and covered a spectral range from 596 nm to 612 nm; the ``CCFE" spectrometer used a 1200 l/mm grating (with a spectral resolution of 0.12 nm) and covered a spectral range from 611 nm to 631 nm ; and the ``Dibs" spectrometer used a 1800 l/mm grating (with a spectral resolution of 0.07 nm) and covered a range from 595 nm to 626 nm. Unfortunately, the ``CCFE" spectrometer did not provide sufficient spectral resolution to accurately pick out many of the Fulcher band lines, in contrast to the ``York" and ``Dibs" spectrometers. This limited high resolution coverage in the lower divertor to the $\nu = 0$ and a part of the $\nu = 1$ vibrational bands via the ``York" spectrometer. Therefore, an analysis of the vibrational distribution was only performed for the upper divertor using the ``Dibs'' spectrometer, whereas the rotational distribution could be analysed for both divertor chambers.\\

%% file: Sections/3.Results_rotational.tex
\section{Results}\label{sec:results}

\subsubsection*{Methodology of inferring the rotational and vibrational distribution of $\text{D}_{2}$ in the MAST-U divertor}

In this section, aspects of the methodology specific to this study are outlined; along with the MAST-U results.\\

For details regarding the $D_2$ Fulcher band, its rotational and vibrational structure, and how these may be determined from Fulcher band spectroscopy, the reader is referred to the appendix.\\

\begin{figure}[hbt!]
    \centering
    \includegraphics[width=0.7\columnwidth]{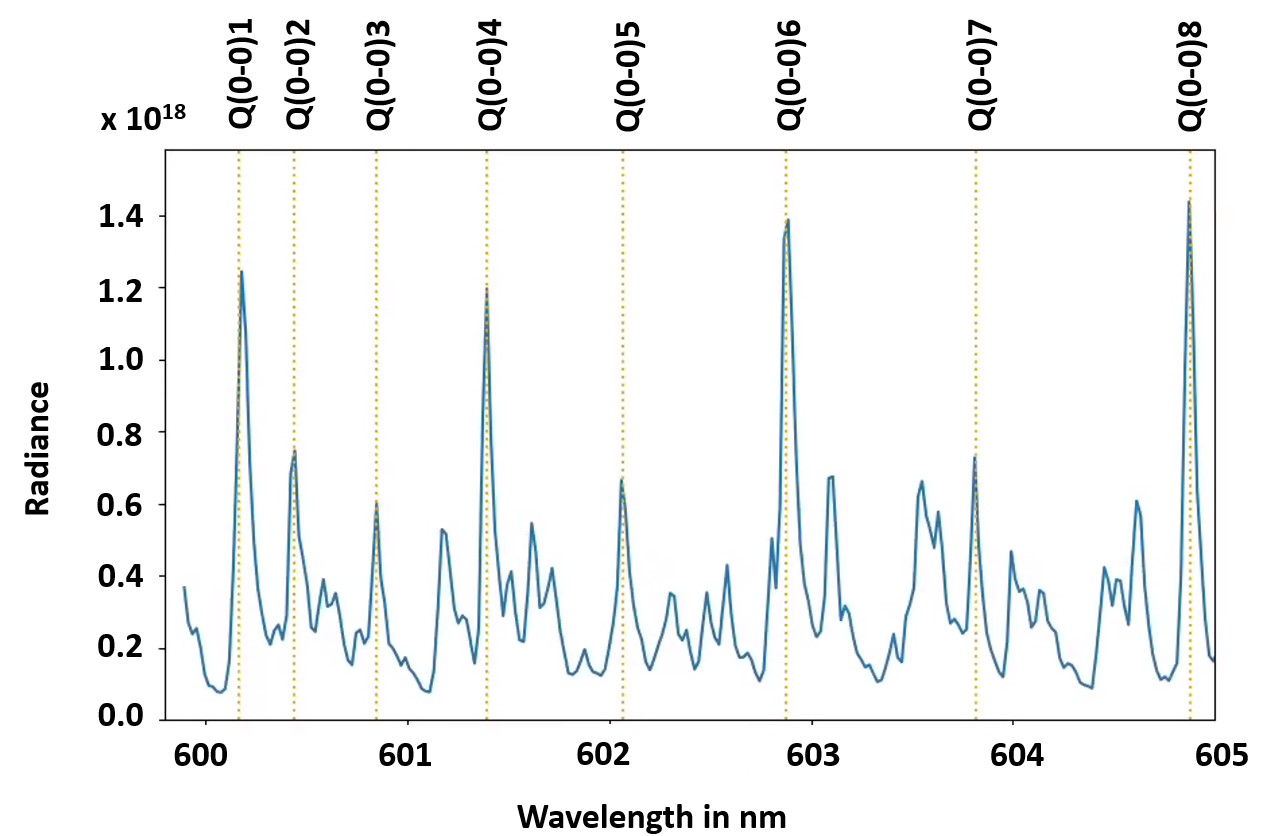}
        \vspace*{0.3cm}
        \caption{\it \footnotesize Example of MAST-U high resolution Fulcher band spectroscopy (data from shot 45244, Dibs spectrometer, line of sight 11, time=500 ms.  The dotted lines labelled Q($\nu^{'}$-$\nu^{''}$)J are the published Q-branch lines taken from \cite{Lavrov2011NewTransitions}.}
        \label{fig:Fulcher_graph}
\end{figure}

A section of a high resolution spectrum obtained from the upper divertor spectrometer (``Dibs") is shown in figure \ref{fig:Fulcher_graph} as an example.  This spectrometer covers all the Fulcher band transitions listed in figure \ref{wavelengths} in the appendix.  Due to the uncertainty in the reproducibility of the wavelength calibration throughout the campaign (1-2 pixels), as well as contaminating impurity transitions, a semi-manual approach was developed to identify the transitions, which was aided by the Boltzmann nature of the rotational distributions. \\ 

Only the relative intensity of the various Fulcher transitions is required to infer the rotational and vibrational distributions. We found that the most reliable method for determining the relative intensities is to use the amplitude of the observed transitions ($H^{'}_{''}$) rather than integrating the peak, since the wings of the spectral lines are often contaminated by other transitions (figure \ref{fig:Fulcher_graph}). The absolute intensity is obtained by integrating the spectral line, which can be approximated by its amplitude times a function representing its shape which is normalised to the peak amplitude: $I^{'}_{''} = H^{'}_{''} \int  f_\lambda (\lambda) d\lambda$. Since all the monitored $D_2$ Fulcher transitions are fully dominated by their instrumental broadening, meaning that $f_\lambda (\lambda)$ is the same for every Fulcher transition, the relative intensities are approximately proportional to their relative amplitudes $H^{'}_{''}$.  If the rotational states form a Boltzmann distribution, then we can write:\\

\begin{equation}
    \mathrm{ln} \left(\frac{H^{'}_{''}\lambda^3}{g_{a.s.}(2J+1)} \right)=\dfrac{-hcBJ(J+1)}{kT_{\textrm{rot}}}+const.
\end{equation}
\\
where $H^{'}_{''}$ replaces $I^{'}_{''}$ in equation \ref{ln_graph} derived in the appendix on page \pageref{sec:theory_rot}.  $J$ is the rotational quantum number;  $2J+1$, and $g_{a.s.}$ account for degeneracy and nuclear statistics respectively; and $B$ is the rotational constant of the molecule.  This allows evaluation of a rotational temperature $T_{\textrm{rot}}$.\\

The experimentally obtained amplitudes are corrected for the plasma background emission (e.g. Bremsstrahlung).  One disadvantage, however,  of replacing the relative intensities with the relative amplitudes is that the sensitivity of the relative intensities to the signal to noise ratio is increased, which drives most of the uncertainty. Additionally, the precise location of a spectral line, with respect to discrete nature of the pixels on the spectroscopy camera, leads to uncertainties in $f_\lambda (\lambda)$ between different transitions, leading to additional uncertainties in $H^{'}_{''}$. Considering the Lorentzian-like nature of the instrumental function with a full-width-half-maximum of roughly 2.5 pixels, this is expected to lead to an uncertainty of 15 \%. If significant overlap with other $D_2$ Fulcher lines or other contaminating transitions occurs, the amplitude is influenced by the contaminating line. Hence, only transitions are used where the closest contaminating line is at least 0.04 nm away. Assuming two spectral lines are 0.04 nm apart with equal intensity, this would introduce an error of 20 \% in the estimation of the amplitude. Due to the various uncertainties, the uncertainty in $H^{'}_{''}$ is at least 25 \%, regardless of the signal to noise ratio.\\

The identification of potentially significant overlapping transitions is aided using the tables provided by \citeauthor{Lavrov2011NewTransitions} \cite{Lavrov2011NewTransitions} together with an inspection of the spectral line shapes. In particular, the $g^3\Sigma^+_g\xrightarrow{}c^3\Pi_u$ transitions interfered with the analysed Q-branch.  The lines impacted by contamination are disregarded from the analysis and the retained transitions were:
\begin{itemize}
    \item Q(0-0): J = 2, 3, 4, 6, 7, 8, 9, 12, 13, 14, 15
    \item Q(1-1): J = 5, 6, 7, 9, 11, 14
    \item Q(2-2): J = 4, 6, 7, 8, 9
    \item Q(3-3): J = 2, 3, 4, 5, 7
\end{itemize}
\label{table: lines}

\subsubsection*{Inferring the $\text{D}_{2}$ rotational temperature in The MAST-U divertor}

\begin{figure}    
     \centering
     \begin{subfigure}[b]{0.48\textwidth}
         \centering
         \includegraphics[width=\textwidth]{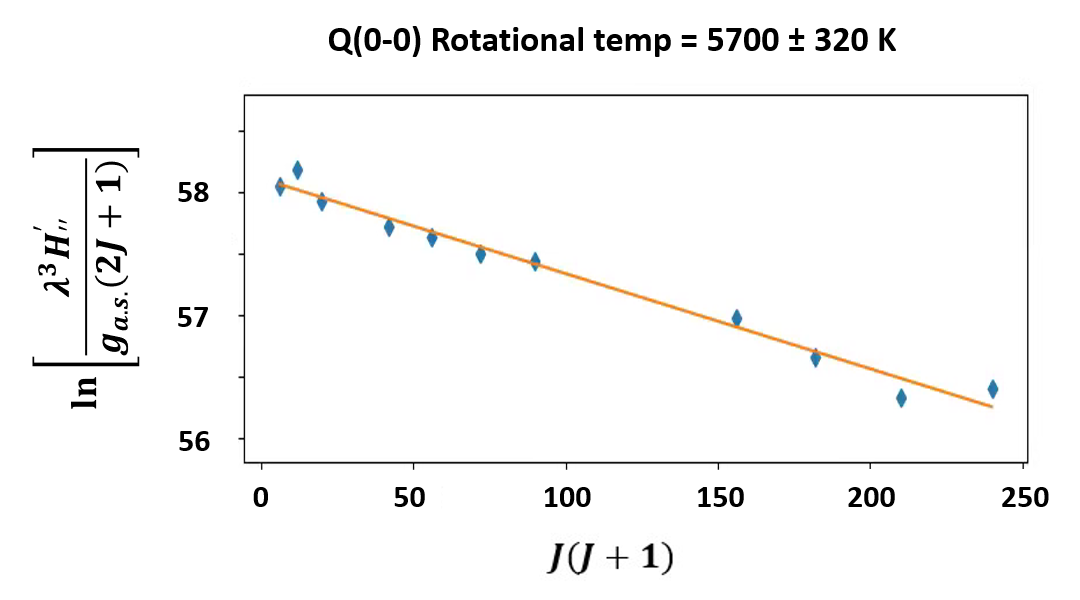}  
         \caption{}
         \label{fig:Q00}
     \end{subfigure}
     \hfill
     \begin{subfigure}[b]{0.48\textwidth}
         \centering
         \includegraphics[width=\textwidth]{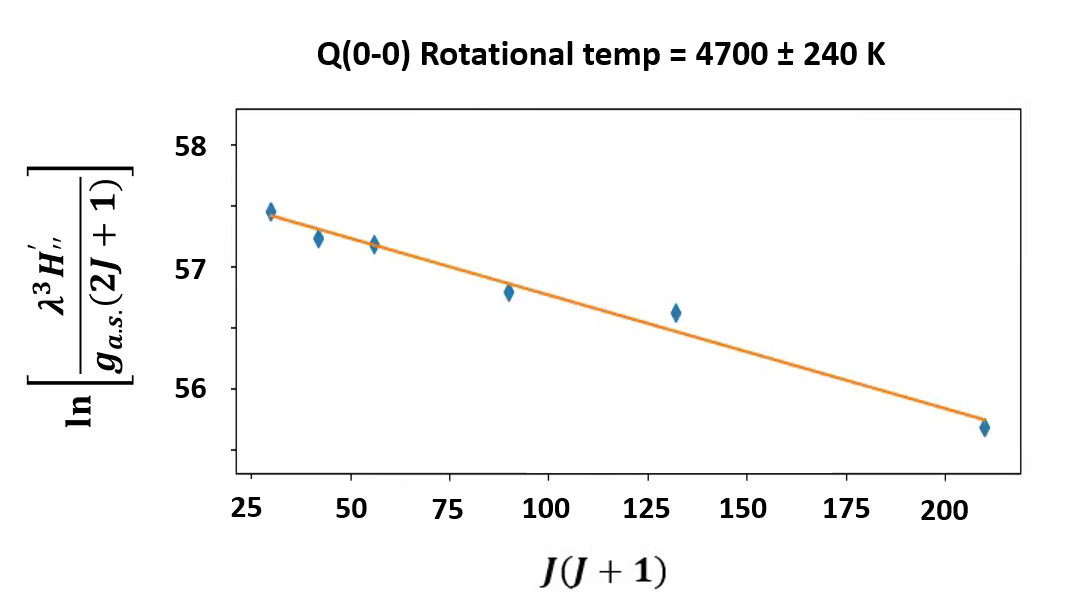}
         \caption{}
         \label{fig:Q11}
     \end{subfigure}   
     \vspace*{0.3cm}
    \caption{\it \footnotesize  MAST-U shot 45244, Upper divertor, line of sight 8, time = 500 ms.  Boltzmann plots of the Q branch transitions.  Plots (a) and (b) show the plots for the Q(0-0) and Q(1-1) respectively.}
        \label{fig:Samples}
\end{figure}

Figure \ref{fig:Samples} shows some examples of the rotational distribution of the upper Fulcher states ($d^3\Pi^-_u$) using upper divertor spectroscopy data. If this distribution follows a linear trend, it means that the rotational distribution follows a Boltzmann relationship and a rotational temperature can be obtained from the linear fit as shown \footnote{Note the rotational temperature has already been mapped to the ground state rotational temperature.}. At sufficient signal to noise levels, the rotational distribution showed good agreements to a Boltzmann scaling for Q(0-0) and Q(1-1) and it was, therefore, possible to extract rotational temperatures from these bands with relatively low uncertainty using a Theil-Sen method.  The uncertainties can be seen as representing the goodness of the Boltzmann fit.  The linear fits for Q(2-2) and Q(3-3) (not shown) have higher uncertainty due to increased signal to noise ratio and may not follow a Boltzmann distribution as clearly as for Q(0-0) and Q(1-1).  There is also a general decrease in the goodness of fit as the rotational temperature increases.  This may be connected to the small Boltzmann gradients getting even closer to zero (see equation \ref{ln_graph}).\\

\subsubsection*{Evolution of rotational temperature and overall Fulcher emission}
\begin{figure}[hbt!]
    \centering
    \includegraphics[width=1\columnwidth]{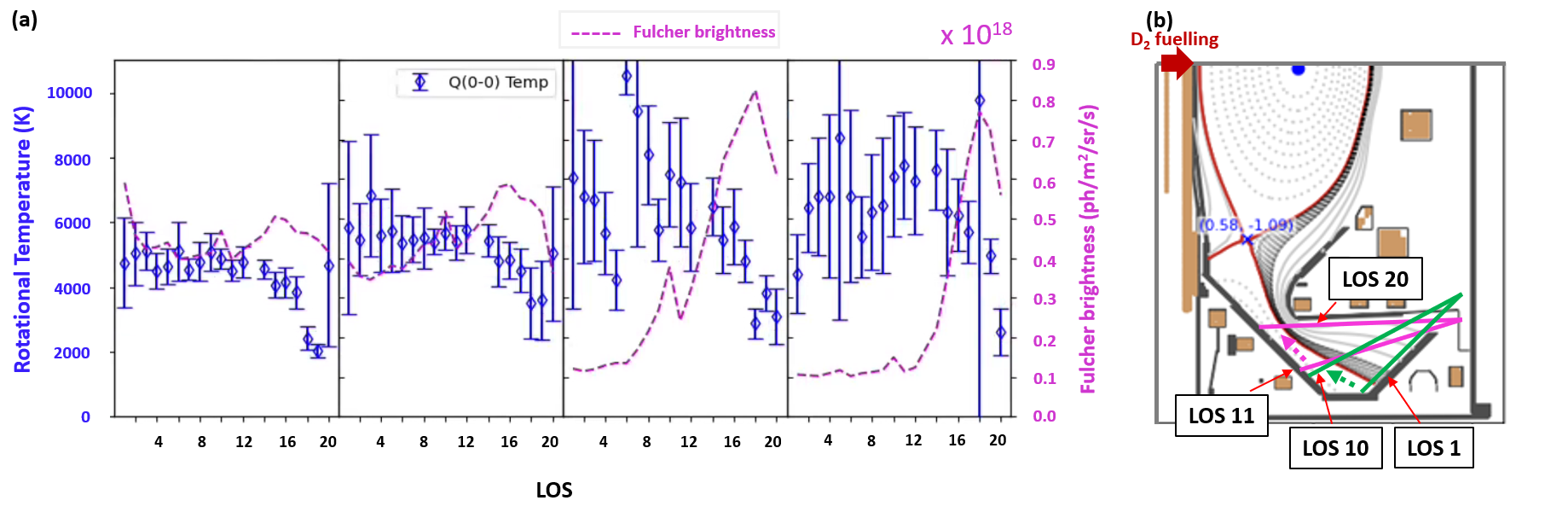}
        \caption{\it \footnotesize MAST-U discharge 45244 Super-X divertor.  (a) Evolution of Fulcher brightness and rotational temperature across the lower divertor (York system).  (b) Lines of sight 1 to 20.}
        \label{fig: York45244}
\end{figure}

\begin{figure}[hbt!]
    \centering
    \includegraphics[width=1\columnwidth]{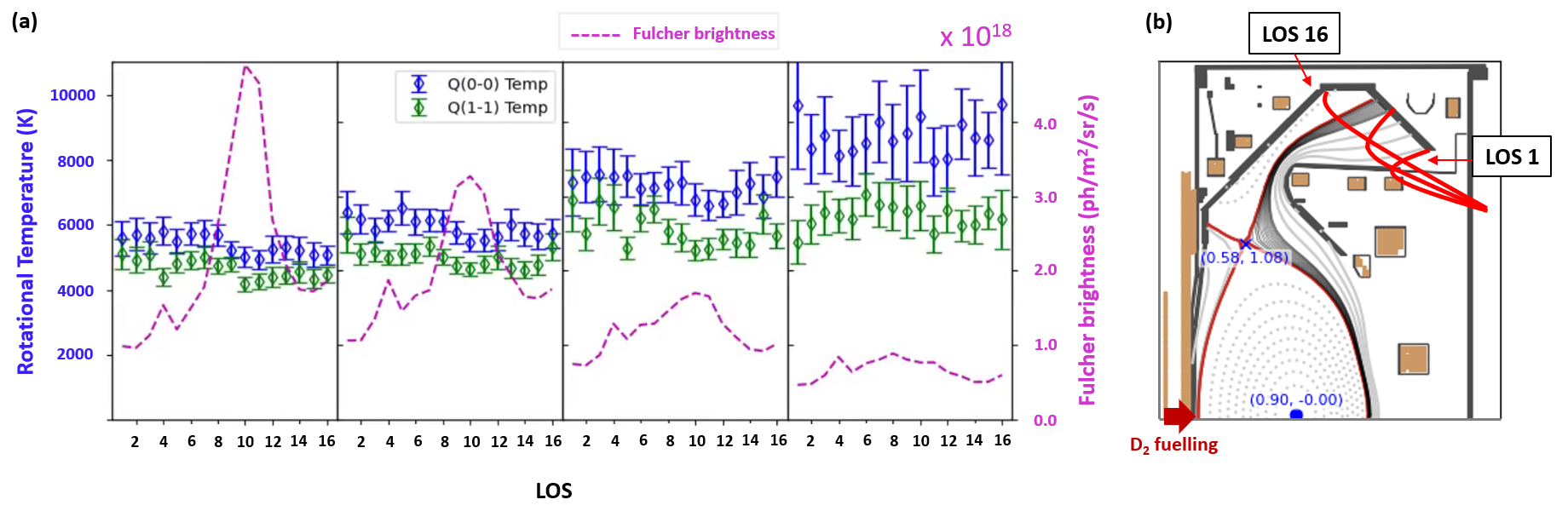}
        \caption{\it \footnotesize MAST-U discharge 45244 Super-X divertor.  (a) Evolution of Fulcher brightness and rotational temperature across the upper divertor (Dibs system), first array.  (b) Lines of sight 1 to 16.}
        \label{fig: 45244Dibs_1to16}
\end{figure}

\begin{figure}[hbt!]
    \centering
    \includegraphics[width=1\columnwidth]{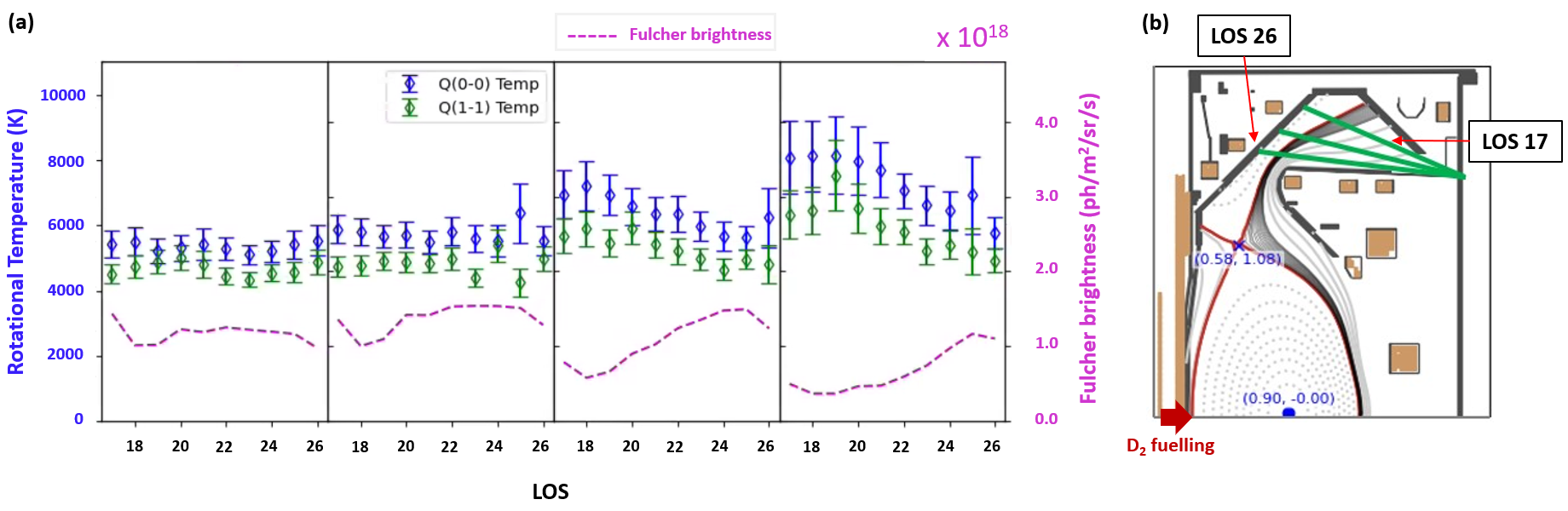}
        \caption{\it \footnotesize MAST-U discharge 45244 Super-X divertor.  (a) Evolution of Fulcher brightness and rotational temperature across the upper divertor (Dibs system), second array.  (b) Lines of sight 17 to 26.}
        \label{fig: 45244Dibs_17to26}
\end{figure}

\begin{figure}[hbt!]
    \centering
    \includegraphics[width=1\columnwidth]{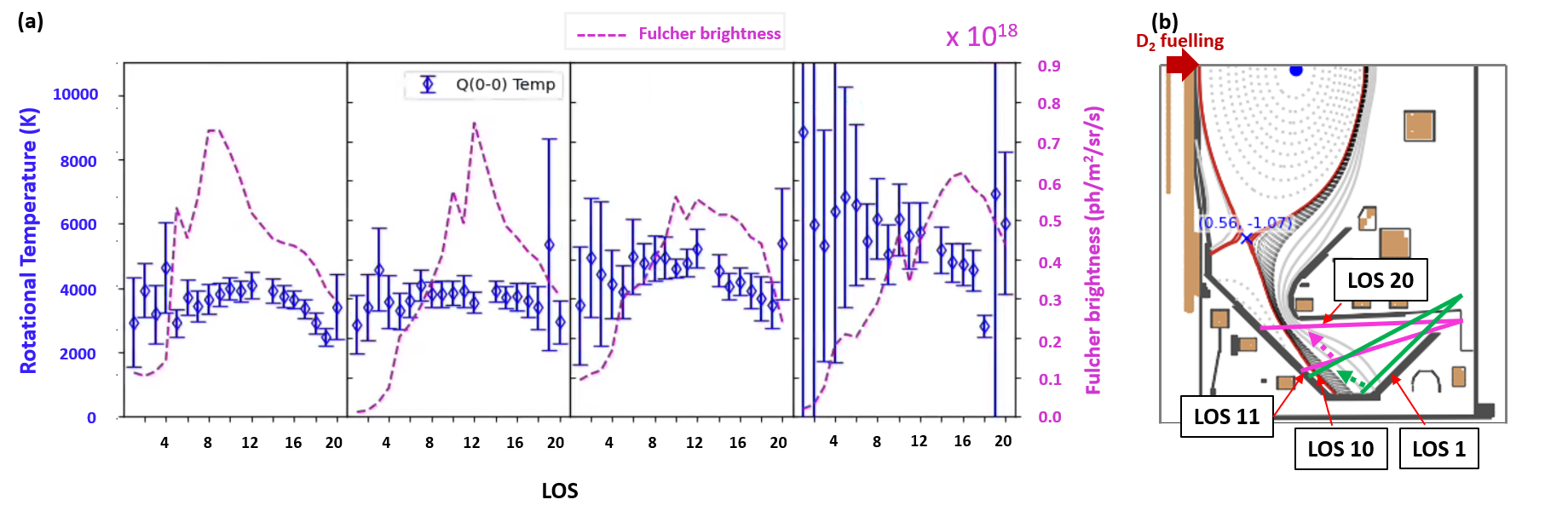}
    \vspace{-0.5cm} 
        \caption{\it \footnotesize MAST-U discharge 45068 elongated conventional divertor.  (a) Evolution of Fulcher brightness and rotational temperature across the lower divertor (York system).  (b) Lines of sight 1 to 20.}
        \label{fig: York45068}
\end{figure}

\begin    {figure}[hbt!]
    \centering
    \includegraphics[width=1\columnwidth]{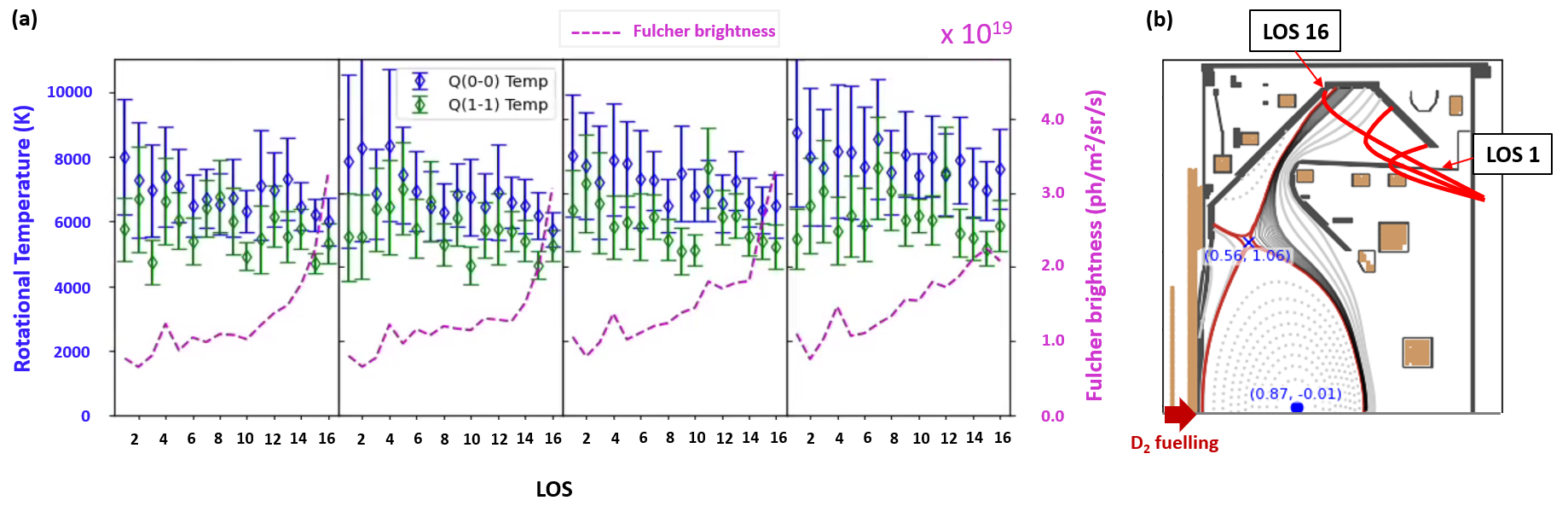}
        \caption{\it \footnotesize MAST-U discharge 45068 elongated conventional divertor.  (a) Evolution of Fulcher brightness and rotational temperature across the upper divertor (Dibs system), first array.  (b) Lines of sight 1 to 16.}
        \label{fig: Dibs45068_1to16}
\end{figure}

\begin{figure}[hbt!]
    \centering
    \includegraphics[width=1\columnwidth]{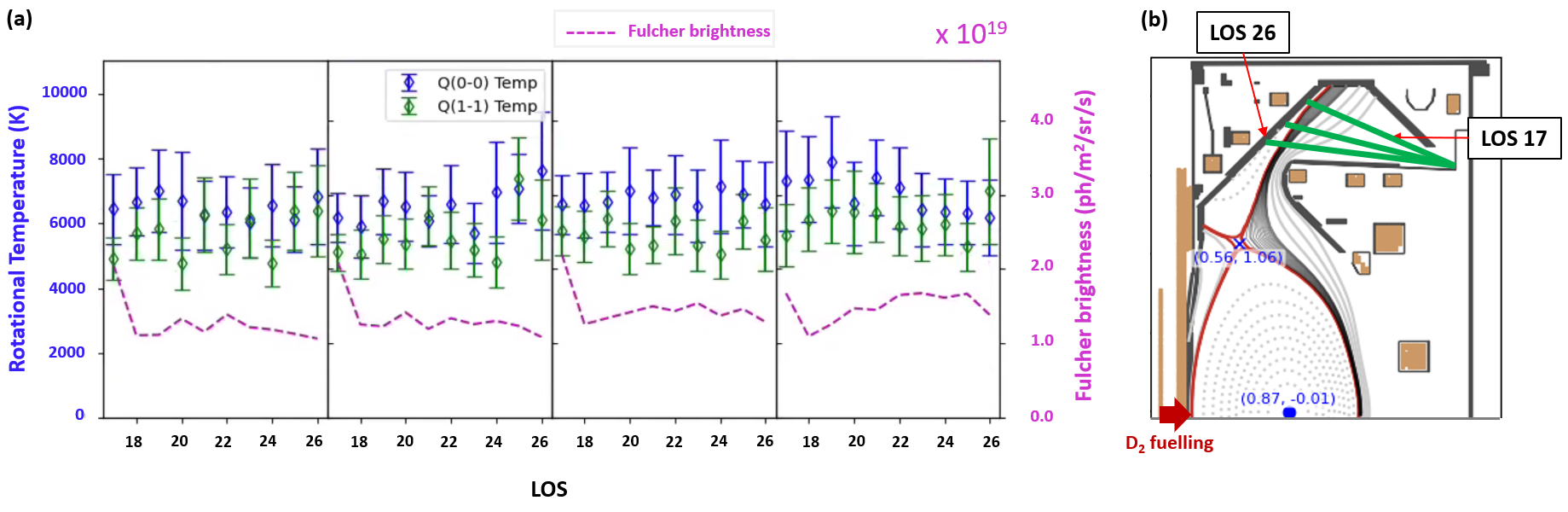}
        \caption{\it \footnotesize MAST-U discharge 45068 elongated conventional divertor.  (a) Evolution of Fulcher brightness and rotational temperature across the lower divertor (Dibs system), second array.  (b) Lines of sight 17 to 26.}
        \label{fig: Dibs45068_17to26}
\end{figure}

Figure \ref{fig: York45244} shows the evolution of the rotational temperature with Greenwald fraction (figure \ref{one}b) during shot 45244 (Super-X divertor configuration) in the lower divertor and figures \ref{fig: 45244Dibs_1to16} and \ref{fig: 45244Dibs_17to26} show the evolution in the upper divertor.  The upper divertor results are divided into two diagrams to depict the observations from the two separate viewing fans (see figure \ref{Ds}b).  Figures \ref{fig: York45068},  \ref{fig: Dibs45068_1to16} and \ref{fig: Dibs45068_17to26} show the same evolutions for shot 45068 (elongated conventional divertor configuration). \\

\begin{figure}[hbt!]
    \raggedright
    \includegraphics[width=1\columnwidth]{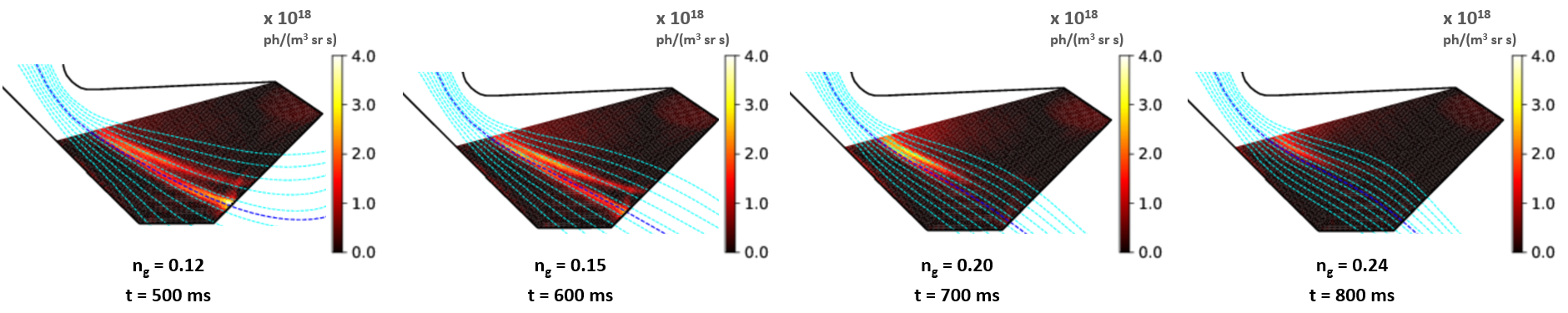} 
        \caption{\it \footnotesize Discharge 45244 (Super-X). Spectrally integrated $D_2$ Fulcher emission maps in 2D at different times, obtained by inverting the MWI measurements.}
        \label{fig: 45244MWI}
\end{figure}

\begin{figure}[hbt!]
    \raggedright
    \includegraphics[width=1\columnwidth]{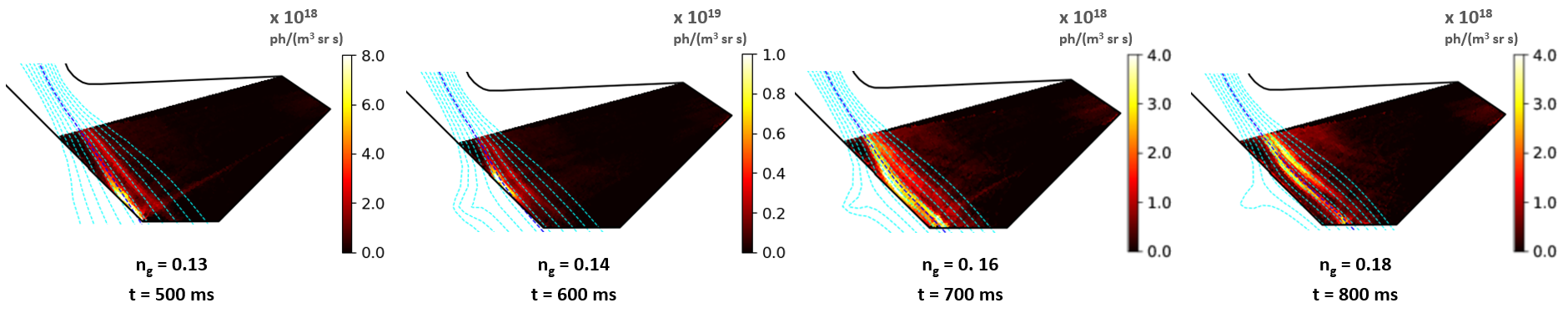}    
        \caption{\it \footnotesize Discharge 45068 (Elongated conventional). Lower divertor Fulcher band emission evolution with Greenwald fraction, obtained from inverting data from the MWI diagnostic \cite{Wijkamp2023}. }
        \label{fig: 45068MWI}
\end{figure}

Spectrally integrated $D_2$ Fulcher emission (spectrally integrated between 598-605 nm) in the lower divertor was captured by the multi-wavelength imaging (MWI) system \cite{Wijkamp2023}.  Its inversions are shown in figures \ref{fig: 45244MWI} and \ref{fig: 45068MWI} for the SXD and ECD fuelling scans respectively, and show the evolution of the 2D Fulcher emissivity profile in the divertor chamber with time (and thus Greenwald fraction). As detachment progresses, the downstream-end of the 2D Fulcher emissivity detaches from the target and moves upstream, which is a proxy for the movement of the ionisation front \cite{Verhaegh2022SpectroscopicDivertor,Verhaegh2021AConditions,Wijkamp2023}.\\

The following should be noted:
\begin{itemize}
    \item The upper divertor poloidal fan (lines of sight 17 to 26) roughly mirrors to the lower divertor lines of sight 9 to 20 as can be seen from inspection of figures \ref{Ds} (a) and (b).
    \item The uncertainty in the absolute brightness of the upper divertor spectrometer is significantly larger than in the lower divertor. Additionally, the upper divertor spectrometer (``Dibs") captures a larger wavelength range than the lower divertor spectrometers, and, in the case of the first array, integration path lengths of the lines of sight through the divertor plasma are longer due to its tangential nature. \textit{Therefore, the absolute Fulcher emission brightnesses cannot be compared between the lower and upper divertor and the upper divertor brightnesses should be considered in a relative sense}.
    \item The ``Dibs" spectrometer was used with an electron-multiplication gain of 10 for discharge 45244 (the Super-X case) and without electron-multiplication gain for discharge 45068 (the elongated conventional divertor case). This has reduced the signal to noise level to such a degree (figures \ref{fig: Dibs45068_1to16} and \ref{fig: Dibs45068_17to26}) that no quantitative rotational temperatures could be inferred. 
\end{itemize}

Figures \ref{fig: 45244Dibs_1to16} and \ref{fig: 45244Dibs_17to26} clearly show a rise in rotational temperature across most of the divertor as detachment occurs and deepens during the Super-X discharge 45244. The ground state rotational temperature rises from below 6000 K at the start of the Super-X at 500 ms and a Greenwald fraction of 0.12, to 8000 K or 9000 K at 800 ms and a Greenwald fraction of 0.24 .  This occurs fairly consistently in all lines of sight except those near to the divertor chamber entrance (21 to 26 in the second ``Dibs" array (see figure \ref{fig: 45244Dibs_17to26})) where the temperature rise is significantly smaller.  The increase in the rotational temperature seems to be correlated with the reduction of the $D_2$ Fulcher band brightness and thus the movement of the Fulcher emission region / ionisation front, as can be observed from the MWI inversions (figure \ref{fig: 45244MWI}). \\

A similar conclusion can be drawn from the lower divertor results shown in figure \ref{fig: York45244}.  The rotational temperature in the lower divertor appears to rise from below 5000 K to 7000 K or 8000 K near the target in the detached region below the ionisation source. Once again the temperature rise is small near the chamber entrance.\\

For the elongated conventional divertor (shot 45068 - see figures \ref{fig: 45068MWI} to \ref{fig: Dibs45068_17to26}) a similar result is observed in the lower divertor.  The rotational temperature rise at the centre of the divertor chamber elevates from around 4000 K to around 6000 K for a change in Greenwald fraction from 0.13 to 0.18. \\

%% file: Sections/4.Results_vibrational.tex
\subsubsection*{The vibrational distribution of $\text{D}_{2}$ in The MAST-U divertor}

\begin{figure}[hbt!]
    \centering
    \includegraphics[width=0.6\columnwidth]{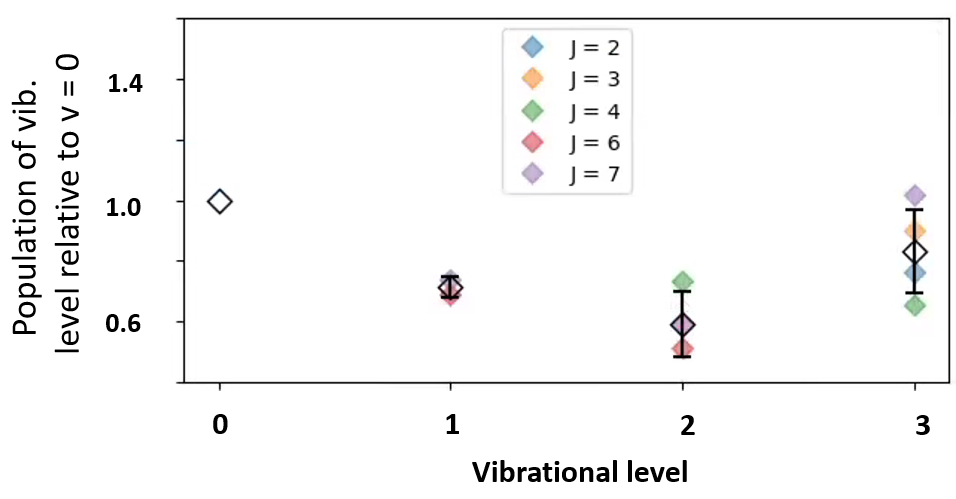}
        \vspace*{0.3cm}
        \caption{\it \footnotesize Typical composite plot showing the vibrational distribution of the upper Fulcher state for the Super-X divertor shot 45244 LOS = 12, time = 600 ms.  The J values represent the rotational quantum numbers of the Q-branch lines available in both the $\nu$ = 0 band and at least one other $\nu >$ 0 band.  The median population is shown by the black diamonds with associated error bars.}
        \label{fig:vib_act}
\end{figure}

Although obtaining the vibrational distribution is more challenging due to considerable overlap of the various vibrational bands, an estimation of relative populations of the zeroth to third vibrational levels in the upper Fulcher state is possible using parts \ref{1st} to \ref{3rd} of the procedure outlined in the appendix on page \pageref{vib_proc}. Due to this type of interference, only certain Q-branch lines were available in each vibrational band (see page \pageref{table: lines}) and barely any transitions preserving a particular rotational number were available in all four vibrational bands.  Therefore, a composite plot was made that compares the population of any Q-branch transition in a vibrational band $\nu > 0$ to band $\nu = 0$ if that transition was available in both bands. This procedure was carried out for the Super-X discharge 45244 and a typical composite plot showing a vibrational distribution is shown in figure \ref{fig:vib_act}. Based on the composite result, an estimate for the relative population and its uncertainty is obtained. \\

A simple Franck-Condon analysis (mapping) for inferring the vibrational distribution of the upper Fulcher state, based on the assumption of a Boltzmann distribution in the ground state, is described in the appendix (see page \pageref{vib}). Figure \ref{fig:vib_theor} in the appendix shows the form of the expected distribution for various ground state vibrational temperatures.  Clearly, the vibrational distribution shown in figure \ref{fig:vib_act} deviates significantly from those calculated in figure \ref{fig:vib_theor}.  This suggests that the ground state distribution may not be distributed according to a Boltzmann distribution.\\

\begin{figure}[hbt!]
    \centering
    \includegraphics[width=1\columnwidth]{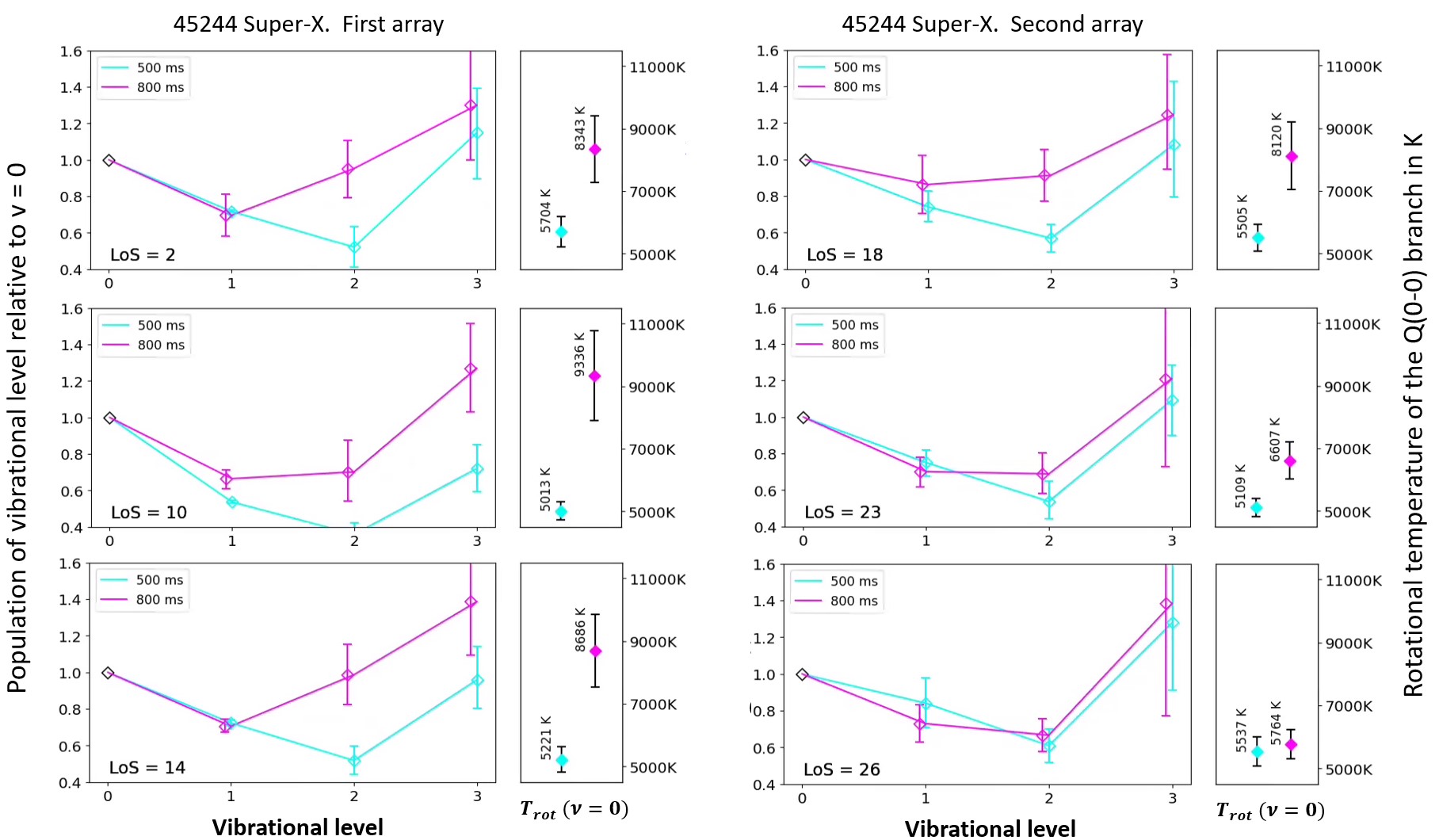}
        \vspace*{0.3cm}
        \caption{\it \footnotesize The vibrational distributions of the first four vibrational bands for various lines of sight in the ``Dibs" arrays at 500 ms and 800 ms.  The small boxes show the ground state rotational temperatures obtained from the Q(0-0) branch for the same lines of sight (the cyan diamond showing the temperature at 500 ms and the magenta diamond at 800 ms).}
        \label{fig:full_vib}
\end{figure}

The temporal and spatial evolution of the vibrational distribution in the MAST-U Super-X divertor during a density ramp discharge is shown in figure \ref{fig:full_vib}. We find that, as detachment occurs and the Fulcher emission moves away from the target (figure \ref{fig: 45244MWI}), the measured vibrational distribution changes in the cold, detached, region near the target - whilst this does not occur near the divertor entrance where the plasma is hotter. As the vibrational distribution changes in the detached region, it deviates further from that expected of a Boltzmann distribution. More precisely, the $\nu = 2,3$ levels increase with respect to the $\nu = 0$ level, whilst the $\nu = 1$ relative population remains constant. This change in vibrational distribution seems to occur simultaneously with a change in rotational temperature, as indicated in figure \ref{fig:full_vib} and the correlation between the two is further investigated in section \ref{sec:discussion}.


%% file: Sections/5.Discussion.tex
\section{Discussion}\label{sec:discussion}

\subsection*{Correlation between rotational temperature and vibrational distribution}
\label{sec:corr_Trot_vib}

\begin{figure}[hbt!]
    \centering
    \includegraphics[width=1\columnwidth]{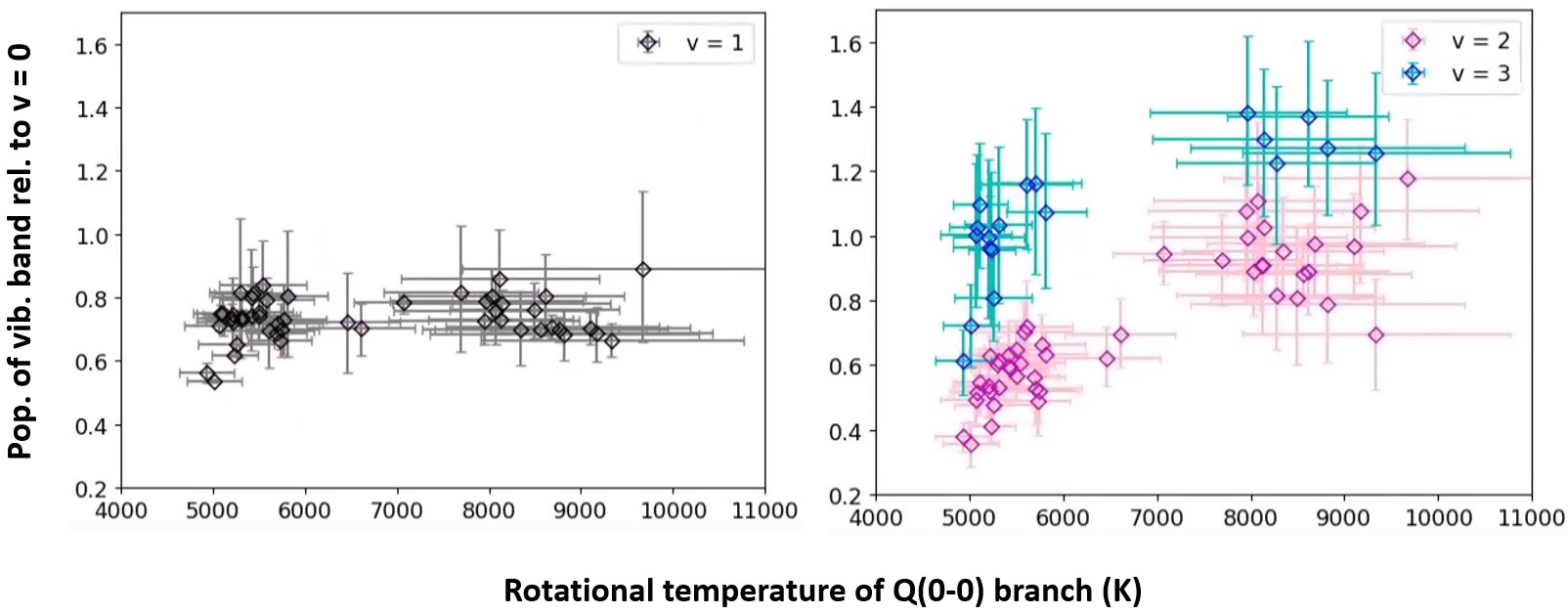}
        \vspace{0.3cm}
        \caption{\it \footnotesize Scatter plots showing the correlation between the populations of the $\nu = 1$ (left) and $\nu = 2,3$ (right) vibrational levels (relative to the $\nu = 0$) with the ground state \textit{rotational} temperature of the Q(0-0) branch.}
        \label{fig: scatter vib}
\end{figure}

In section \ref{sec:results}, we have observed that the deviation between the vibrational distribution and that expected from a Boltzmann (in the ground state) becomes larger in the detached region. At the same time, the inferred rotational temperatures are increased in the detached region (figure \ref{fig:full_vib}). In this section, we study the correlation between the rotational temperature, which is generally assumed to be a measure of the gas temperature, and the vibrational distribution in more detail.  Figure \ref{fig: scatter vib} shows scatter plots of rotational temperature of the ground state Q(0-0) branch against $\nu=1$, $\nu=2$ and $\nu = 3$ population (relative to $\nu=0$). This shows that (in the $D_2$ Fulcher excited state) the populations of the $\nu=2$ and $\nu=3$ vibrational levels relative to the $\nu=0$ level are roughly proportional to the \textit{rotational} temperature of the molecules, which is not the case for the $\nu=1$ band.\\

The strong, positive correlation between $\nu=2,3$ with $T_{\textrm{rot}}$ shows that, as the gas temperature is increased, there is an increase in the relative population of $\nu=2,3$ in the $D_2$ upper Fulcher state.  It is also the case that the deviation between a distribution expected from a Boltzmann distribution (in the ground state, mapped to the $D_2$ Fulcher state with a Franck-Condon approach) and that observed becomes larger.\\

Our results (section \ref{sec:results}) indicate that the increase in the gas temperature occurs spatially below the ionisation source. Hydrogen Balmer line analysis in similar discharges, has shown that a build-up of molecules and neutral atoms occurs below the ionisation region \cite{Verhaegh2022SpectroscopicDivertor,Verhaegh2023b}. The increase of molecular density, results in plasma-chemistry interactions with vibrationally excited molecules, leading to molecular ions (such as $D_2^+$) that react with the plasma, leading to Molecular Activated Recombination (MAR) and Dissociation (MAD) \cite{Verhaegh2022SpectroscopicDivertor,Verhaegh2023b}. The appearance of MAR in the Super-X and elongated divertors occurs downstream of the ionisation region and builds up as the plasma grows more deeply detached. At the same time, we find that the rotational temperature increases as the plasma becomes more deeply detached (section \ref{sec:results}). Therefore, the MAR/MAD strength, gas temperature and the level to which the measured vibrational distribution deviates from that expected of a Boltzmann distribution (in the ground state) all seem correlated. 

\subsection*{Vibrational excitation mechanisms and deviations from a Boltzmann distribution}
\label{sec:vibr_excit}

The observation that a Franck-Condon analysis suggests deviation of the vibrational distribution from that of a Boltzmann in the ground state could mean that: 1. the simplified Franck-Condon analysis used is insufficient to explain the vibrational distribution in the upper state. (The Franck-Condon approach may be invalid, for example, if there is vibration-rotation interaction \cite{Farley2011}, which is the case when inter-molecular interactions are relevant); and/or 2. that reactions occur that cause the vibrational distribution in the ground state to deviate from a Boltzmann distribution.\\

The correlation between the appearance of MAR and the increased deviation between the measured vibrational distribution and that expected of a Boltzmann distribution in the ground state lends weight to the 2nd point - i.e. that plasma-chemistry interactions alter the vibrational distribution beyond that expected of a Boltzmann in the ground state.

There are a variety of processes which may contribute to such deviation:
\begin{itemize}
    \item Reactions between the plasma and the molecules can lead to the (de)population of specific vibrational states: for example, molecular charge exchange and dissociative attachment (which can generate molecular ions) is more likely for higher vibrational levels and can thus depopulate the high-vibrational distributions \cite{Ichihara2000,Verhaegh2021AConditions,Verhaegh2022SpectroscopicDivertor,Reiter2018}.
    \item Conversely, repeated excitation and de-excitation of one of the singlet electronic states, incurring a shift in vibrational quantum number, can pump the higher vibrational levels in the ground state \cite{Chandra2023,Holm2022}.
    \item Vibrationally excited molecules have sufficiently large mean free paths below the ionisation region in detached divertors for transport effects to be important \cite{Holm2022}. Molecules may have an initial vibrational distribution when being released from the wall \cite{Wischmeier2005,Verhaegh2023a}, which - combined with transport - means the distribution would possess a wall material dependency.
    \item Under certain conditions of high molecular density, vibrationally excited molecules can interact with each other leading to vibrational-vibrational energy exchange \cite{Krasheninnikov1996PlasmaDetachment}. In this interaction, one molecule loses a quantum of vibrational energy and another gains a quantum of vibrational energy. The molecular anharmonicy of the vibrational potentials can mean that the kinetic temperature can drive the vibrational distribution into the so-called Treanor distribution which overpopulates higher vibrational states.  The Treanor distribution is characterised by a vibrational temperature $T_{\nu}$ and a gas temperature $T_0$:
    \bigskip
    \begin{equation}
    N(\nu)=N_0\,exp\left(\dfrac{-\hbar \omega \nu}{T_{\nu}}+ \dfrac{\hbar \omega \chi_e \nu(\nu+1)}{T_0}\right)
    \end{equation}
    
    where $\chi_e=0.0196$ is the anharmonicity parameter for ${D}_{2}$ \cite{Fridman2008PlasmaChemistry, Krasheninnikov1996PlasmaDetachment, Smirnov2001PhysicsGases}.
    \item Analogously to the case for a rotational distribution (see appendix), a multi-temperature component of the vibrational distribution may exist. This could be because different interactions can drive different vibrational temperatures; as well as due to the spectroscopic line-of-sight integrating through parts of the plasma where Fulcher emission at a different vibrational temperature occurs.
\end{itemize}

\begin{figure}[hbt!]
    \centering
    \includegraphics[width=0.8\columnwidth]{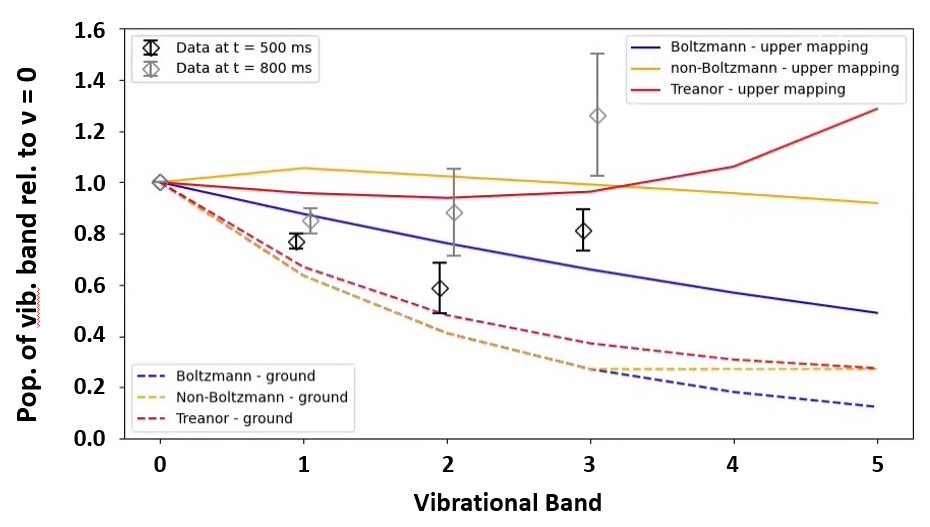}
        \vspace*{0.3cm}
        \caption{\it \footnotesize Various ground state vibrational distributions mapped onto the upper Fulcher state using a Franck-Condon approach.  The blue lines represents a Boltzmann distribution at a vibrational temperature of 9500 K; the orange lines represents a distribution that is Boltzmann (again at 9500 K) for the first four bands and is then flat; the red lines shows a Treanor distribution with a vibrational temperature of 9500 K and a gas temperature of 2500 K.  The black and grey diamonds show actual data for the first four vibrational bands at the start and end of the Super-X in shot 45244.}
        \label{fig:Tr_vs_Boltz}
\end{figure}

Figure \ref{fig:Tr_vs_Boltz} shows the mapping of three different ground state distributions (dotted lines) to the upper Fulcher state (solid lines) using a Franck-Condon approach (as described in the appendix on page \pageref{vib}) and compares these mappings to real data from Super-X shot 45244 at 500 ms and 800 ms (black and grey diamonds).  The blue lines show the mapping for a Boltzmann distribution at a vibrational temperature of 9500 K.  The orange lines show the mapping for a distribution which follows a Boltzmann (again at 9500 K) for the first four bands and then becomes flat.  This distribution approximates experimentally observed plasmas in low temperature plasma experiments and simulations \cite{Capitelli2006VibrationalAspects, Mosbach2005PopulationPlasma} where higher vibrational states are seen to become overpopulated.  The red lines show the mapping for a Treanor distribution (with vibrational temperature at 9500 K and gas temperature at 2500 K).  The value of 9500 K for the vibrational temperature was chosen as a temperature which allowed a reasonable mapping in all cases.  Two things are clear:  firstly none of the distributions map well on to the data; and, secondly, relatively small differences in the ground state populations have a pronounced effect on the upper state under Franck-Condon mapping.\\

It is noted that the Treanor distribution (see figure \ref{fig:Tr_vs_Boltz}), which arises from vibrational-vibrational energy exchange, bears some resemblance to the actual vibrational distributions observed in the MAST-U divertor in the sense that the higher vibrational levels can have an elevated population. Vibrational-vibrational exchange occurs at higher molecular density, which is significantly increased in the cold, detached, region where the elevation of the higher vibrational levels (in the Fulcher state) is enhanced. Additionally, the tightly baffled nature of the MAST-U divertor chamber significantly elevates the molecular density.  However, once mapped to the upper Fulcher state, it shows significant deviation to the measured distribution.\\

At this stage, it is unclear what mechanisms are driving the vibrational distribution on MAST-U. This requires further studies where the vibrational distribution is modelled and compared against experimental measurements.

\subsection*{Relevance and implications of our results}

Our results are the first measurements of the rotational and vibrational distribution of $D_2$ for an alternative divertor configuration with tight divertor baffling. In the MAST-U Super-X divertor, deeper levels of detachment were obtained \cite{Verhaegh2022SpectroscopicDivertor,Verhaegh2023b}, in which plasma-molecular chemistry involving MAR and MAD had a stronger impact than on other devices. These results show a correlation between detachment, the increase of the rotational temperature of the molecules and a deviation between the measured vibrational distribution and that expected from a Boltzmann distribution of the ground state, with elevated levels of $\nu=2,3$ in the $D_2$ Fulcher state. In this section we discuss the relevance and implications of these results, as well as how it compares against results from other devices and the relevance for reactors.\\

Analysis of the $D_2$ Fulcher band emisson to infer rotational and vibrational distributions has been performed on a range of different devices in the past, including tokamaks (JET \cite{Sergienko2013MolecularJET}, DIII-D \cite{Hollmann2006SpectroscopicEdge}, ASDEX-Upgrade \cite{Fantz2001}, JT60-U \cite{Kubo2005}), as well as linear devices such as Magnum-PSI \cite{Akkermans2020}. Generally, a Boltzmann distribution is found for the rotational distribution across the various devices. A large range of different rotational temperatures are quoted in literature, ranging from less than 1000 K (mainly linear devices) to 10000 K and higher \cite{Hollmann2006SpectroscopicEdge}. In various devices an increase of the rotational temperature with the electron density is observed, particularly JET \cite{Sergienko2013MolecularJET}.  \citeauthor{Hollmann2006SpectroscopicEdge} shows similar rotational temperatures, at the same electron density, for DIII-D, TEXTOR (limiter, no divertor) as well as for a linear device, PISCES \cite{Hollmann2006SpectroscopicEdge}. The spread in the various reported rotational temperatures may depend on whether the molecular density arises predominantly from recycling or from fuelling.\\

Our MAST-U results - in the novel, tightly baffled, Super-X divertor chamber, are consistent with these historic results. It shows a clear Boltzmann trend of the rotational distribution and the rotational temperatures obtained are consistent with previous findings. $n_e \sim 10^{19} \mathrm{\,m^{-3}}$ is expected for the studied discharges in the divertor, which does not change greatly during a density ramp due to the large detached operational space \cite{Verhaegh2022SpectroscopicDivertor}. At such low densities, much lower rotational temperatures are expected based on DIII-D, PISCES and TEXTOR \cite{Hollmann2006SpectroscopicEdge} results. The MAST-U Super-X results suggest that the increase of rotational temperature is not necessarily associated with divertor electron densities, but with the level of detachment (in diverted tokamaks). Higher rotational temperatures, if the rotational temperature is a proxy for the kinetic temperature of the molecules, can be expected in the tightly baffled MAST-U Super-X chamber as the molecules arise from continuous recycling in which additional energy and momentum transfer can occur from the plasma to the molecules. Potentially, the high rotational temperatures on MAST-U imply that not all the emission from the upper Fulcher state arises from electron-impact excitation (e.g. other processes, such as recombination, could result in electronic excited molecules), which requires further research. \\

In most divertors it was possible to infer a vibrational temperature from the vibrational distribution. However, JET-ILW results have reported an overpopulation in $\nu=2,3$ \cite{Sergienko2013MolecularJET}, when compared to expectations based on previous, carbon wall, measurements. Although the vibrational distribution measurements on MAST-U are roughly in line with measurements on other devices, in terms of the observed relative populations of $\nu=1,2,3$, the increase of the relative distribution of $\nu=2,3$ with increasing $T_{\textrm{rot}}$ has not been previously observed.\\

The observed differences in the MAST-U Super-X divertor may be caused by certain processes that are predominantly relevant in the Super-X divertor. The large distance of the ionisation source from the target results in a large volume with relatively long molecular mean free paths, meaning that molecular transport could play a significant role. The high molecular densities may imply that molecule-molecule collisions, such as vibrational-vibrational exchange, could be significant - which would be less likely on other devices. This has implications for both alternative divertor concepts as well as tightly baffled divertors. This means that ADC specific research is required, in addition to that on more `conventional' divertors, to validate the models used for the vibrational distribution in plasma-edge physics. As an initial step, 0D collisional-radiative models can be used to see if they can match the observed vibrational distribution and which processes play a role in this. This information can then be included in plasma-edge simulations, both in a resolved and unresolved setup to test whether transport of vibrationally excited molecules and plasma-wall interactions can have a significant impact.\\

The plasma conditions will be significantly different in ADC reactor designs than on MAST-U, operating at higher power, higher densities and with a metallic wall. This will change the plasma-wall interactions and result in smaller mean free paths. However, if such a design is deeply detached, there can still be a significant region below the ionisation front in which the divertor is detached, in which the various described processes could be relevant. Plasma-edge modelling, using the outcomes of the above validation steps in a transport-unresolved study, could be used to probe the sensitivity of ADC reactor designs to such interactions.

%% file: Sections/6.Summary.tex
\section{Conclusion} \label{summary}

Our work provides the first analysis of the rotational and vibrational distribution of $D_2$ molecules in the tightly baffled, deeply detached MAST-U Super-X divertor using $\text{D}_{2}$ Fulcher band emission spectroscopy during a density ramp discharge where the level of detachment is scanned. The rotational distribution follows a Boltzmann distribution very clearly for the first two vibrational bands; and, with greater uncertainty, for the 3rd and 4th bands.  In contrast, the vibrational distribution of the upper Fulcher band (measured from $\nu = 0$ to $\nu = 3$) deviates from that expected if an assumed Boltzmann distribution in the ground state is projected into the $D_2$ upper Fulcher state using a Franck-Condon analysis.  The $\nu=2,3$ populations, relative to $\nu=0$, in the $D_2$ Fulcher state show a significant overpopulation compared to that expected from a ground state Boltzmann distribution.\\

At detachment onset, the ground state rotational temperature, which is often assumed to be an indicator of the kinetic energy of the molecules, is observed to increase from below 6000 K to 9000 K near the target. As the core density is scanned and detachment proceeds, the region of elevated rotational temperatures expands upstream, following the ionisation front. In this region, strong signatures of plasma-chemistry interactions leading to Molecular Activated Recombination / Dissociation are observed based on hydrogen Balmer line emission spectroscopy. At the same time, the measured vibrational distribution changes significantly in the cold, detached, region: both the $\nu = 2$ and $\nu = 3$ vibrational bands (in the $D_2$ upper Fulcher state) receive a population boost relative to the $\nu = 0$ band. The change of the vibrational distribution during detachment is larger than that expected of a (ground state) Boltzmann distribution. The relative population of the $\nu=2,3$ levels is strongly correlated with the $\textit{rotational}$ temperature of the ground state Q(0-0) branch; whereas the relative population of the $\nu=1$ level appears to be independent of the detachment state or rotational temperature.\\

Both of these observations suggest a kinetic link between the plasma and the $\text{D}_{2}$ molecules; which changes across the entire detached region with larger changes occurring in deeper detached conditions. This work provides the tools and measurements required to provide a deeper understanding of this kinetic link and its relation to detachment; which will be used to compare against models in the future. \\

\newpage

%% file: Sections/Appendix.tex
\section*{Appendix}
\subsection*{The Fulcher band of $\text{D}_{2}$}\label{sec:theory_app}

The molecular hydrogen Fulcher band is a band of visible lines between 600 nm and 650 nm generated by de-excitations of electronically excited states of the hydrogen molecule from $d^3\Pi^-_u\xrightarrow{}a^3\Sigma^+_g$ and can be prominent in divertor plasmas.  These excited states are ro-vibronic and transitions of the form $n^{'}\nu^{'}J^{'}\xrightarrow{}n^{''}\nu^{''}J^{''}$ must be considered where $n^{'}, \nu^{'}, J^{'}$ are the electronic, vibrational and rotational quantum numbers of the upper state; and $n^{''}, \nu^{''}, J^{''}$ are the corresponding quantum numbers of the lower state.\\

The so-called ``Q-branch" transitions are relatively bright transitions in which both the vibrational and rotational quantum numbers are preserved.  By observing the molecular Fulcher band spectra with a high resolution spectrometer, it is then possible to utilise the intensities of such transitions to determine information about the rotational distribution in a particular vibrational band of the upper electronic state; and the vibrational distribution of bands within the upper electronic state.  Information can then potentially be inferred about the rotational and vibrational distributions in the ground state.  If such a distribution follows a Boltzmann function, it can be characterised with a certain temperature.\\

\begin{figure}[hbt!]
    \centering
    \includegraphics[width=0.6\columnwidth]{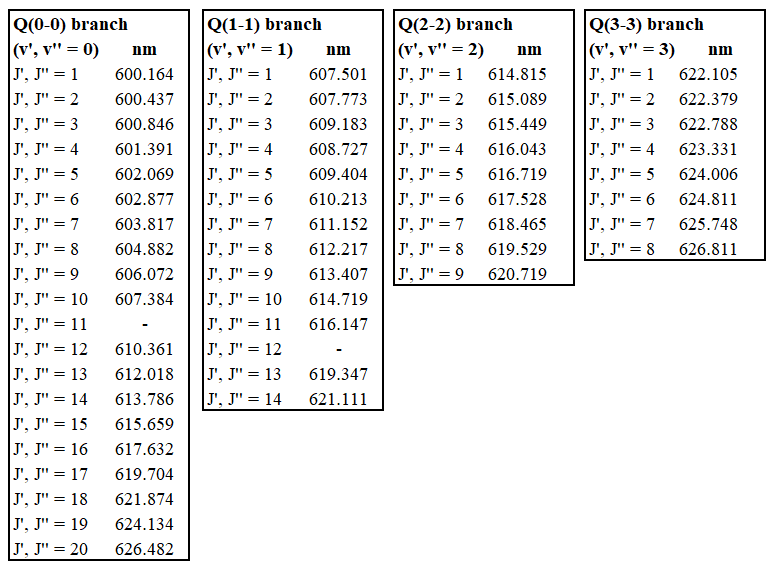}
        \vspace*{0.3cm}
        \caption{\it \footnotesize Table showing the $d^3\Pi^-_u\xrightarrow{}a^3\Sigma^+_g$ Q-branch transitions accessible through the MAST-U high resolution divertor spectroscopy. $\nu^{'}$ and $\nu^{''}$ represent the upper and lower vibrational states, and $J^{'}$ and $J^{''}$ the upper and lower rotational states.  Data obtained from \citeauthor{Lavrov2011NewTransitions} \cite{Lavrov2011NewTransitions}.}
        \label{wavelengths}
\end{figure}

Figure \ref{wavelengths} shows the $d^3\Pi^-_u\xrightarrow{}a^3\Sigma^+_g$ Q-branch transitions which were accessible from high resolution MAST-U divertor spectroscopy in the first experimental campaign, using the documentation provided by \citeauthor{Lavrov2011NewTransitions} \cite{Lavrov2011NewTransitions}.  In this paper a Q-branch transition is described in the usual way using the terminology $Q(\nu^{'}-\nu^{''})J$ where $J=J^{'}=J^{''}$.\\

\subsection*{The Rotational Distribution of $\text{D}_{2}$}\label{sec:theory_rot}
If the diatomic molecule $\text{D}_2$ is considered as a rigid rotator, solutions of the appropriate Schr{\"o}dinger equation are possible only for specific energy values:\\
\begin{equation}
    E=\dfrac{h^2J(J+1)}{8\pi^2I}
\end{equation}
where $J$ is the rotational quantum number which has integer values 0, 1, 2, ... and $I$ is the moment of inertia of the molecule.\\

The wave number of a photon emitted or absorbed during the transition between two rotational energy states $E^{'}$ and $E^{''}$ is given by:\\
\begin{equation}
    \dfrac{1}{\lambda}=\dfrac{E^{'}}{hc}-\dfrac{E^{''}}{hc}
\end{equation}
and it is convenient to define the so-called ``rotational term":\\
\begin{equation}
    F(J)=\dfrac{E}{hc}=\dfrac{h}{8\pi^2cI}J(J+1).
\end{equation}
If $B$ is defined ``rotational constant" which is unique for any particular rotator:\\
\begin{equation}
    B=\dfrac{h}{8\pi^2cI}
\end{equation}
then the rotational term becomes:
\begin{equation}
    F(J)=BJ(J+1).
\end{equation}

We can now consider the a thermal distribution of such rotational states according to the Boltzmann factor:\\
\begin{equation}
    \text{e}^{-E/kT}\propto \text{e}^{-hcBJ(J+1)/kT_{rot}},
\end{equation}
where $T_{rot}$ is a ``rotational temperature" defining the distribution.  The vibrational energy can be ignored since we are considering rotational states within the same vibrational band, and hence this constant addition factors out of the exponential.\\

Quantum theory tells us that for each rotational state of total angular momentum $J$, there is a $2J+1$ degeneracy.  We can, therefore, write that the molecular density $n_J$ in a particular rotational level $J$ is such that:\\
\begin{equation}
    n_J\propto (2J+1)\text{e}^{-hcBJ(J+1)/kT_{rot}}
\end{equation}

In order to relate the specific spectral lines to this distribution, it is necessary to consider the intensities of the lines.  The emissivity for a given transition ($\epsilon^{'}_{''}$ in $ph/m^3/sr/s$) will be proportional to the density of molecules in the upper state, $n_J$,  and to the Einstein coefficient of the transition $A^{'}_{''}\propto \dfrac{1}{\lambda^3}$. However, a spectrometer measures a line intensity, which is integrated along the line of sight: $I^{'}_{''} = \int_L \epsilon^{'}_{''} (x) dx$. Assuming that the emissivity is constant along the line of sight, for a path length $\Delta L$, this would imply that: $I^{'}_{''} = \Delta L \epsilon^{'}_{''}$. In this case, we can write:\\
\begin{equation}\label{Boltz}
    I^{'}_{''}\propto \Delta L n_J\,A^{'}_{''}\,\propto \,\dfrac{1}{\lambda^3}\,g_{a.s.}\,(2J+1)\,\text{e}^{-hcBJ(J+1)/kT_{rot}}
\end{equation}
where $g_{a.s.}$ is a factor incorporated to account for nuclear statistics.  Nuclear spin permutations introduce the following weightings:\\
\begin{equation}
    \dfrac{\text{Number of ways of achieving odd}\,J}{\text{Number of ways of achieving even}\,J}=\dfrac{(S+1)}{S} \text{for nuclei with half integral spin; and}
\end{equation}
\begin{equation}
    \dfrac{\text{Number of ways of achieving odd}\,J}{\text{Number of ways of achieving even}\,J}=\dfrac{S}{(S+1)} \text{for nuclei with integral spin.}
\end{equation}

The deuterium molecule has nuclei with $S=1$ and therefore the ratio is $1:2$.  This ratio determines the factor $g_{a.s.}$ where $g_{a.s.}=g_{a.}$ and $g_{a.s.}=g_{s.}$ represent the weightings for odd and even values of $J$ respectively.  In the case of $D_2$, therefore $g_{a.}=1$ and $g_{s.}=2$.\\

\subsubsection*{Rotational Temperature}
By taking the natural log of equation \ref{Boltz}, it can be tested whether a measured rotational distribution follows a Boltzmann distribution as a linear relationship is expected (equation \ref{ln_graph}). Using the slope of such a linear relationship, the rotational temperature can be inferred \cite{Majstorovic2007RotationalDischarge, Qing1996DiagnosticsRegime, Briefi2017DeterminationTransition, Gavare2010PlasmaBand}:\\
\begin{equation}\label{ln_graph}
    ln \left(\frac{I^{'}_{''}\lambda^3}{g_{a.s.}(2J+1)} \right)=\dfrac{-hcBJ(J+1)}{kT_{rot}}+const.  
\end{equation}

If a rotational temperature is obtained in this way, i.e. from the distribution of the upper state, then it is clear that it pertains to the distribution in the excited state.  The assumption that is usually made \textit{in the case of a low density plasma} is that this excited distribution of rotational states is a mapping of the distribution in the ground state.  As is discussed in \cite{Bruggeman2014GasReview} this assumption that the excited distribution maps a ground state distribution is based on the following:
\begin{itemize}
    \item that the lifetime of the ground-state is long enough for thermalisation to occur through collisions;
    \item that the lifetime of the excited state is short in comparison with the rotational relaxation time through collisions; and
    \item that during excitation there is very little change in rotational quantum number.
\end{itemize}
If this assumption is valid, as may be expected in a low density plasma, then it can be said that:\\
\begin{equation}\label{ratio}
    \dfrac{B^{0}}{T^{0}_{rot}}=\dfrac{B^{*}}{T^{*}_{rot}},
\end{equation}
and we can simply use the rotational constant for the ground state $B^{0}$ in a plot of equation \ref{ln_graph} and extract the ground state rotational temperature directly.\\

For the states under consideration for $D_2$, $B^{0}=30.4\,\si{cm^{-1}}$ and $B^{*}=15.2\,\si{cm^{-1}}$\\

Importantly, we can also take the first bullet-point in the above list to imply that (in a low density plasma) the rotational temperature of the ground state ought to represent the kinetic gas temperature. It should also be noted, though, that another state of affairs may lead to a Boltzmann distribution of the rotational states in the excited state.  In a high density/highly collisional plasma, a distribution which may be non-Boltzmann for some reason in the ground state, may experience enough collisionality in the lifetime of the excited state to become thermalised and therefore exhibit a Boltzmann distribution of rotational states - i.e. for the opposite reason to the second bullet-point. It seems logical that in this situation, the rotational temperature of the excited state ought to represent the gas temperature.\\

Finally we note, as also detailed in \cite{Bruggeman2014GasReview}, that it is possible for a plasma to exhibit non-Boltzmann rotational distributions when a reaction mechanism leads to molecules that are excited to a specific rotational state in the ground state, combined with the lifetime of the excited state not being long enough for thermalisation by collisions. A common example of a non-Boltzmann distribution is the so-called ``two-temperature" Boltzmann distribution which is often reported in laboratory plasma experiments \cite{Briefi2020APlasmas, Vankan2004HighPlasmas, Bruggeman2014GasReview, Ishihara2021Ro-vibrationalSpectroscopy}.  In these cases, higher rotational quantum numbers are overpopulated with respect to a Boltzmann distribution.  Meanwhile, thermalisation via rotational relaxation is inefficient for higher rotational quantum number (this is expected since the cross-sections for rotational energy transfer diminish rapidly with increasing J).  In such cases, it is typically assumed that the temperature of the lower J Boltzmann fit is the best estimate of the gas temperature (once corrected via equation \ref{ratio}). Multiple-temperature Boltzmann distributions may also be caused by chordal integration effects, for example if $D_2$ Fulcher emission occurs at two different locations along the line of sight with two different rotational temperatures.\\

\subsection*{The vibrational Distribution of $\text{D}_{2}$} \label{vib}
Unlike the rotational quantum number, vibrational quantum number is much less likely to be preserved during electronic excitation or radiative de-excitation. Therefore, in contrast to the rotational temperature inferences, the excited distribution no longer maps directly to the ground state distribution. Instead, for vibrational transitions, the excitation from the electronic ground state $X^1\Sigma^+_g$ to the excited state $d^3\Pi^-_u$ must be considered in a vibrationally resolved manner using Franck-Condon factors, as must the Fulcher band transitions from $d^3\Pi^-_u\xrightarrow{}a^3\Sigma^+_g$.  A simplified approach to infer the ground state vibrational temperature can be used which assumes a Bolzmann distribution in the ground state and then maps this to the upper Fulcher state using Franck-Condon factors \cite{Fantz1998SpectroscopicMolecules, Briefi2020APlasmas}.  In its simplest form, this approximation assumes instantaneous decay from the upper state and a constant electron dipole transition moment and proceeds as follows:
\label{vib_proc}
\begin{enumerate}
    \item \label{1st} Select Q-branch lines which are available (with little interference) in all vibrational bands and use their summed intensities as a proxy for the intensity of the entire vibrational band.  (Note that the line intensities must be adjusted so as to account for the rotational temperature of their band).
    \item \label{2nd} Normalise these intensities to the $\nu$ = 0 band intensity.
    \item \label{3rd} Divide by appropriate branching ratios to determine the relative populations of vibrational levels in the upper Fulcher state $d^3\Pi^-_u$.
    \item \label{4th} Assume a Boltzmann distribution of vibrational states in the ground state $X^1\Sigma^+_g$ and normalise to the $\nu=0$ state.
    \item \label{5th} Use Franck-Condon factors to determine the distribution in the excited state $d^3\Pi^-_u$ that would be theoretically produced via electron impact excitation of the ground state distribution.
    \item Vary the vibrational temperature in \ref{4th} and repeat \ref{5th} to fit to the measured distribution in \ref{3rd}.
\end{enumerate}

The Franck-Condon factors between the first 20 vibrational states for transitions $d^3\Pi^-_u\xrightarrow{}a^3\Sigma^+_g$ are available \cite{Fantz2004Franck-CondonIsotopomeres}. 
The branching ratios are shown in table \ref{fig:table}.\\

\begin{table}[h]
    \centering
    \caption{\it \footnotesize The branching ratios for transitions from $d^3\Pi^-_u\xrightarrow{}a^3\Sigma^+_g$ which preserve vibrational quantum number (adapted from \cite{Fantz1998SpectroscopicMolecules}).}\label{fig:table}
    \vspace*{0.3cm}
    \begin{tabular}{|c|c|}
        \hline
        \textbf{D$_{2}$ transition $d^3\Pi^-_u\xrightarrow{}a^3\Sigma^+_g$} & \textbf{Branching ratio}\\
        \hline
        $\nu^{\prime}=0\rightarrow \nu^{\prime\prime}=0$ & 0.871 \\
        $\nu^{\prime}=1\rightarrow \nu^{\prime\prime}=1$ & 0.695 \\
        $\nu^{\prime}=2\rightarrow \nu^{\prime\prime}=2$ & 0.543 \\
        $\nu^{\prime}=3\rightarrow \nu^{\prime\prime}=3$ & 0.412 \\
        \hline
    \end{tabular}
\end{table}

\begin{figure}[hbt!]
    \centering
    \includegraphics[width=0.6\columnwidth]{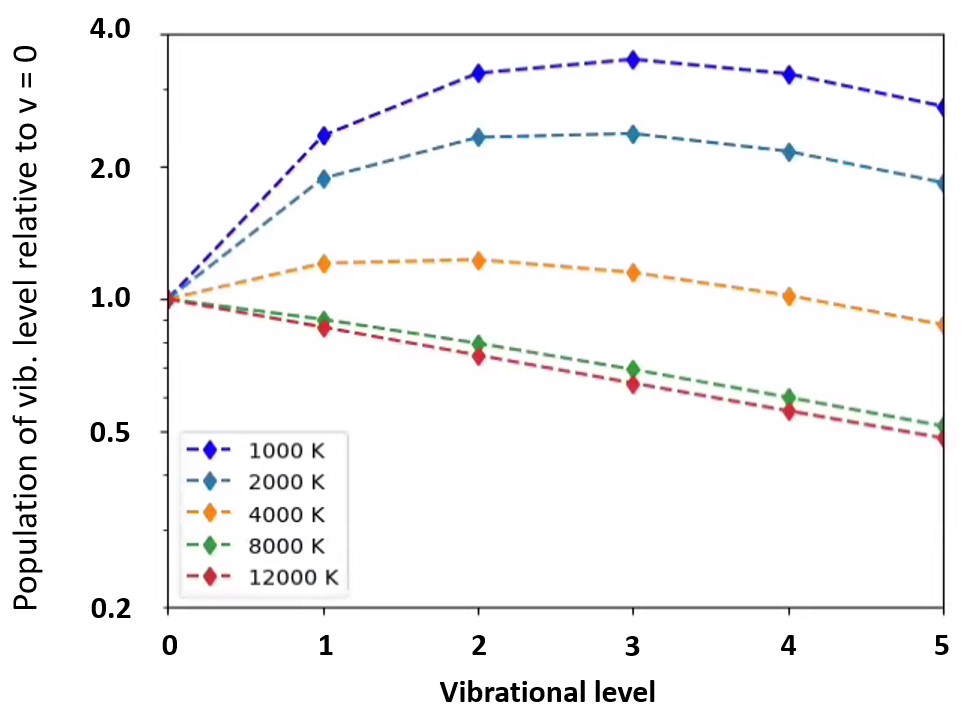}
        \vspace*{0.3cm}
        \caption{\it \footnotesize Theoretical vibrational distributions of the upper Fulcher state for various vibrational temperatures are shown by the dotted lines.}
        \label{fig:vib_theor}
\end{figure}

An example of application of the parts \ref{1st} to \ref{5th} of this process to generate theoretical vibrational distributions of the upper state, assuming the vibrational distribution in the ground state is distributed according to a Boltzmann at a certain temperature, are shown in figure \ref{fig:vib_theor}.\\

